\documentclass[
    reprint,
    preprintnumbers,
    superscriptaddress,
    nofootinbib,
     amsmath,amssymb,
     aps,
     prd,
    floatfix,
    ]{revtex4-2}

\usepackage{graphicx}
\usepackage[dvipsnames]{xcolor} 
\usepackage[utf8]{inputenc}
\usepackage{color}
\usepackage[normalem]{ulem} 
\usepackage{physics}
\usepackage{placeins}
\usepackage{comment}
\usepackage{booktabs, multirow}
\usepackage{enumitem}
\usepackage{comment}
\usepackage{bm}
\usepackage{bigints}
\usepackage{appendix}
\usepackage{enumitem}
\usepackage{comment}
\usepackage{makecell}
\usepackage{array}
\usepackage{orcidlink}
\usepackage{rotating}
\usepackage{cleveref}
\usepackage{verbatim}
\usepackage{cancel}
\usepackage{mathtools}
\usepackage[nolist,nohyperlinks]{acronym}
\usepackage{dsfont}

\hypersetup{
  colorlinks   = true, 
  urlcolor     = orange, 
  linkcolor    = blue, 
  citecolor   = blue 
}
    \allowdisplaybreaks[1]
    \graphicspath{{figures/}}

\newcommand{\potsdam}{Institut f{\"u}r Physik und Astronomie, Universit{\"a}t Potsdam, Haus 28, Karl-Liebknecht-Str. 24/25, 14476, Potsdam, Germany}
\newcommand{\aei}{Max Planck Institute for Gravitational Physics (Albert Einstein Institute), Am M{\"u}hlenberg 1, Potsdam 14476, Germany}
\newcommand{\grasp}{Institute for Gravitational and Subatomic Physics (GRASP), Utrecht University, Princetonplein 1, 3584 CC Utrecht, The Netherlands}
\newcommand{\nikhef}{Nikhef, Science Park 105, 1098 XG Amsterdam, The Netherlands}
\newcommand{\soton}{School of Mathematical Sciences, University of Southampton}

\newcommand{\lonebb}{$\Lambda_{1,\rm \textsc{bbh}}$}
\newcommand{\ltwobb}{$\Lambda_{2,\rm \textsc{bbh}}$}
\newcommand{\lonenb}{$\Lambda_{1,\rm \textsc{nsbh}}$}
\newcommand{\ltwonb}{$\Lambda_{2,\rm \textsc{nsbh}}$}

\definecolor{pumpkin}{rgb}{1.0, 0.46, 0.09}

\definecolor{orchid}{rgb}{0.85, 0.44, 0.84}

\definecolor{deeppink}{rgb}{1.0, 0.08, 0.58}

\definecolor{dukeblue}{rgb}{0.0, 0.0, 0.61}

\definecolor{pastelgreen}{rgb}{0.47, 0.87, 0.47}

\definecolor{smokeytopaz}{rgb}{0.58, 0.25, 0.03}

\definecolor{cyan(process)}{rgb}{0.0, 0.72, 0.92}

\Crefname{equation}{Eq.}{Eqs.}

\begin{document}

\title{A story about a tipsy kangaroo: Reversible jump MCMC for model selection in the analysis of gravitational-wave signals from the coalescence of compact objects}

\author{Anna Puecher~\orcidlink{0000-0003-1357-4348}}
\email{anna.puecher@uni-potsdam.de}
\affiliation{\potsdam}
\author{Tim Dietrich~\orcidlink{0000-0003-2374-307X}}
\affiliation{\potsdam}
\affiliation{\aei}
\author{Hauke Koehn~\orcidlink{0009-0001-5350-7468}}
\affiliation{\potsdam}
\author{Jonathan Gair}
\affiliation{\aei}
\author{Gregory Ashton}
\affiliation{\soton}
\author{Luca Negri~\orcidlink{0009-0001-5468-0721}}
\affiliation{\grasp}
\affiliation{\nikhef}
\author{Anuradha Samajdar~\orcidlink{0000-0002-0857-6018}}
\affiliation{\grasp}
\affiliation{\nikhef}

\date{\today}

\begin{abstract}

Bayesian inference is commonly employed in the analysis of gravitational-wave signals not only to estimate the source parameters, but also for model selection. 
The latter provides insight into the physics of the source and has the potential to inform the direction for future model development. 
Although model comparison is usually performed by analyzing the data separately with different models and comparing the obtained Bayesian evidences, an alternative approach consists in sampling directly over the model itself. 
Here, we present \texttt{t-roo}, a reversible jump Markov chain Monte Carlo sampler capable of performing transdimensional inference on gravitational-wave signals from compact binary coalescences.
Employing \texttt{t-roo}, a single analysis provides simultaneously the model odds ratio and the parameter posteriors for the favored models, hence yielding a potentially substantial computational advantage, particularly when comparing many models or analyzing highly informative data.
\texttt{t-roo} is built on the sampler \texttt{eryn} and is specifically designed to compare models describing different kinds of sources, i.e., binary black hole, binary neutron star, or neutron star-black hole systems, as well as multiple models for the same source class. 
We validate the sampler on a set of injections, finding agreement with the results obtained with the nested sampler \texttt{dynesty}. We then use \texttt{t-roo} to analyze the real events GW190425 and GW230529, for which the system's parameters alone do not provide conclusive evidence of the presence of a neutron star component. 
\texttt{t-roo} can be adapted to any model-comparison scenario, thus providing a valuable tool in particular for next-generation detectors, where analyzing data separately with competing models becomes computationally even more demanding.
\end{abstract}

\maketitle

\section{Introduction}

In the past ten years, a growing number of gravitational wave (GW) detections from different sources has enabled us to study the population and properties of binary systems of compact objects~\citep{LIGOScientific:2026wfs, LIGOScientific:2026ctl}, to place constraints on the nuclear equation of state and cosmological parameters~\citep{LIGOScientific:2017vwq, LIGOScientific:2017ync, LIGOScientific:2017adf, Dietrich:2020efo, LIGOScientific:2026uyd}, and to test the theory of General Relativity~\citep{LIGOScientific:2021sio, LIGOScientific:2026qni, LIGOScientific:2026fcf, LIGOScientific:2026wpt}.
The properties of the emitting system are commonly inferred from GW signals through Bayesian inference, not only to measure the source's parameters, but also to test hypotheses and compare different models in order to determine which one provides the best description of the data.
Bayesian analyses require estimating parameter posteriors or model evidences, a computationally challenging task for the highly-dimensional parameter spaces involved in GW data-analysis. 
Therefore, stochastic sampling methods, such as nested or Markov chain Monte Carlo (MCMC) samplers~\cite{nested, mcmc, hastings}, are usually employed to estimate these quantities. 
Currently, various software exist for this purpose and are regularly employed by the LIGO-Virgo-KAGRA (LVK)~\cite{LIGOScientific:2014pky, VIRGO:2014yos, 10.1093/ptep/ptaa125} collaboration to analyze GW data, e.g., Refs.~\cite{dynesty, Ashton:2021anp, Ashton:2018jfp}.

Adopting different approximants to analyze GW detections from a given type of source can hint at which features of the various models describe the data best, thus providing useful information for the development of future models, as shown in Refs.~\cite{Ashton:2021cub,Puecher:2023rxw} for binary neutron star (BNS) and binary black hole (BBH) detections, respectively.

However, in the context of GW analyses from compact binary coalescences (CBC), the relevant question often becomes what was the nature of the objects in the system from which the signal originated.
Unless a clear electromagnetic counterpart is present, the classification of a GW signal source is usually based on the estimated component masses~\cite{LIGOScientific:2025pvj, alerts_guide}, or, if the signal-to-noise ratio (SNR) is large enough, on the presence of tidal information in the phase. 
Nevertheless, for events such as GW190814~\cite{LIGOScientific:2020zkf}, GW230529~\cite{LIGOScientific:2024elc}, and GW190425~\cite{LIGOScientific:2020aai}, the nature of the components remains uncertain (see, e.g., Refs.~\cite{Barbieri:2020ebt, Dudi:2021abi, Foley:2020kus, Kyutoku:2020xka, Han:2020qmn, Khadkikar:2025awf, Tews:2020ylw, Lopes:2021yga, Lee:2021yyn, Biswas:2020xna, Bombaci:2020vgw, Koehn:2024ape, Janquart:2024ztv, Markin:2025oeo}).
Even for GW170817~\cite{LIGOScientific:2017vwq}, an exact source classification based only on the GW signal would not be possible.

In these situations, analyzing GW data with different models can provide information beyond the mere comparison of component masses, thus helping classify the signal's source.
Gravitational-waveform models incorporate effects specific to the source they aim at describing: BNS models include tidal phase information, alongside its interplay with spin, while the amplitude corrections in neutron star - black hole (NSBH) models describe the different behaviors near merger and the possibility of tidal disruption.

Bayesian statistics provides a natural way to compare different models or hypotheses describing a certain signal.
This is conventionally achieved by performing separate analyses on the same data with the various models investigated, and then comparing evidences to compute the so-called odds ratio.
Alternatively, the model can be considered as one of the parameters to sample over. 
The main advantage of this second approach is that the more informative the data, the more time the sampler will spend in the correct model, reducing the iterations in the disfavored ones and the overall computational cost. 

References~\cite{Ashton:2021cub, Puecher:2023rxw} employed a specific case of this second method, introducing a categorical parameter to describe the model to be used at each sampler iteration. 
However, this approach cannot be applied when comparing models with different parameter-space dimensionality.
This issue arises, for example, when comparing aligned-spin and precessing models, or models describing different sources, since in the case of NS components also tidal effects should be accounted for.

In these cases, when we want to compare models with different dimensionalities, we must resort to reversible jump MCMC (RJMCMC)~\cite{rjmcmc, rjmcmc_sum}.
RJMCMC samplers are commonly employed for LISA (Laser Interferometer Space Antenna) data analysis~\citep{Karnesis:2023ras, Umstatter:2005su, Umstatter:2005jd}, for X-ray binaries analyses~\cite{xray}, and for GW ground-based detectors in the context of modelling a burst (or noise) signal with a flexible number of given basis functions, for example in relation to glitch mitigation~\citep{Cornish:2014kda, Littenberg:2014oda, Tong:2024ipe, Chatziioannou:2021ezd, Cheung:2026myt}.

In this work, we develop, to the best of our knowledge, the first RJMCMC sampler to analyze full GW signals from different CBC sources with state-of-the-art waveform approximants. 
In particular, our sampler, called \texttt{t-roo} (Tipsy kangaROO)\footnote{Regarding the name, \textit{kangaroo} was chosen to recall the \textit{jump} in ``reversible jump", and for consistency with Australian names for samplers such as \textsc{Bilby} and \textsc{Dingo}. The kangaroo is \textit{tipsy} because it not only jumps, but it can also perform \textit{reverse} jumps.}, can analyze GW data employing multiple BBH, NSBH, or BNS models at the same time. 
In the end, the sampler provides parameter posteriors and odds ratios between the different models. 

The main advantages of this approach, compared to performing separate runs and then comparing evidences, are:
\begin{enumerate}[label=(\roman*)]
\item A significant reduction of the ``human overhead" during the analysis, since only one run, with no model-specific settings, is needed.
\item A reduction of the computational cost, especially for informative events. In particular, the computational advantage increases the more models are compared at the same time.
 
\item Although the odds ratio convergence requires some time, \texttt{t-roo} provides the source classification rather quickly, and could therefore be employed in the future also for low-latency purposes.
\end{enumerate}

Our sampler is based on elements from \textsc{Bilby}~\cite{Ashton:2018jfp}, where it relies especially on its well-tested utilities for evaluating the likelihood function for CBC signal analyses, and the \texttt{eryn}~\cite{Karnesis:2023ras} sampling package, from which it inherits the structure of the RJMCMC sampler, including parallel tempering. 

We test \texttt{t-roo} performance on simulated signals for both current and future-generation detectors, such as the Einstein Telescope (ET), and we analyze the real GW events GW190425 and GW230529. 

The paper is structured as follows: in Sec.~\ref{sec:t-roo} we provide an overview of the elements inherited from \texttt{eryn} and \textsc{Bilby}, while in Sec.~\ref{sec:proposals} we illustrate the specific proposals used in \texttt{t-roo}. 
In Sec.~\ref{sec:methods} we describe the analysis methods, and we show the results of a set of analyses on simulated data and on the real events in Sec.~\ref{sec:results}.
We summarize our conclusions in Sec.~\ref{sec:conclusions}. In Appendix~\ref{sec:rjmcmc_theory}, we provide a more detailed and technical description of the working mechanisms of RJMCMC samplers, focusing on the specific application to our case in Appendix~\ref{sec:app_rjmcmc_troo}. We illustrate some possible empirical criteria for convergence in Appendix~\ref{sec:convergence_crit}, and in Appendix~\ref{sec:evidences} we discuss the robustness of the evidence computation by comparing the estimates obtained with different samplers. 

\texttt{t-roo} is publicly available on GitHub at \href{https://github.com/AnnaPuecher/t_roo}{\textsf{https://github.com/AnnaPuecher/t\_roo}}.

\section{t-roo basics}
\label{sec:t-roo}

MCMC sampling algorithms differ from nested ones in that they directly sample the target distribution. 
The main condition that MCMC samplers have to satisfy is the so-called \emph{detailed balance}, which essentially means that the probability of a chain being in a state $x$ and transitioning to a state $x'$ must be the same as being in the state $x'$ and transitioning to $x$.
In the specific implementation of the Metropolis Hastings algorithm~\cite{10.1093/biomet/57.1.97}, at each step, we propose a new point in the parameter space and decide whether to accept it or not, where the exact form of the acceptance probability is what ensures detailed balance. 
RJMCMC is a generalized version of this algorithm, which allows the chain to move also across parameter spaces with different dimensionalities.
A detailed overview of MCMC and RJMCMC samplers is provided in Appendix~\ref{sec:rjmcmc_theory} . 

For CBC analyses, MCMC algorithms are conventionally employed to sample the source posterior $p(\vec{\theta} \mid d, M)$, where $\vec{\theta}$ denotes the parameters of interest, $d$ the detector data, and $M$ the model employed to describe the data~\cite{emcee,Ashton:2021anp}.  
If, instead, we have a set of models $\left\{ M_k\right\}$ and want to establish which one describes the data best, we can either analyze the data separately with the various models and then compare evidences, usually computed with nested samplers, or, with MCMC samplers, we can directly sample the joint probability 
\begin{equation}
    p(\vec{\theta}_k, k \mid d) = p(\vec{\theta}_k \mid k, d) p(k \mid d),
\end{equation}
where $k$ is the model label and $\vec{\theta}_k$ the parameters of model $k$.
Notably, reversible jump MCMC allows us to do this even if the models considered have different dimensions. 
The better a model manages to describe the data, the more time the sampler will spend in that model. 
Therefore, once the sampler is converged, the final probability of a model will be given by 
\begin{equation}
   p_{{\textsc{mod}}_k} \coloneqq  p(k \mid d) =  \frac{n_k}{\sum_k n_k},
\end{equation}
where $n_k$ is the number of samples specifically in the model $k$.

\texttt{t-roo} is based on \texttt{eryn}~\cite{Karnesis:2023ras}, a sampler purposefully designed to tackle Bayesian analyses of complex and multi-dimensional problems, such as the LISA global fit, which is in turn based on the affine invariant sampler \texttt{emcee}~\cite{emcee}. 
In MCMC ensemble samplers, a collection of $N_w$ \emph{walkers} explore the parameter space in parallel, with proposals based on the position of other walkers. 
The key property of affine invariant samplers is that they have the same efficiency in sampling different distributions related by affine transformations. 
This implies that they can rather efficiently sample highly-correlated distributions, which are commonly encountered in GW analysis from CBC sources. 

In the following, we provide a brief explanation of the features that \texttt{t-roo} inherited from \texttt{eryn} and \texttt{Bilby}.

\subsection{Stretch move}
\label{sec:stretch_move}
In Ref.~\cite{goodmanweare}, the authors suggest the so-called \emph{stretch move} as the simplest proposal to update the walkers positions that satisfies both affine invariance and detailed balance.
In a stretch move, the new position of a walker $X_i$ at time $t$ is proposed by moving by a fraction of the distance with respect to another random walker $X_j$ as 
\begin{equation}
    X_{i,{\rm new}} = X_j(t) - z (X_i(t) - X_j(t)),
\end{equation}
with $z$ being a random variable drawn from a distribution $g(z)$.
This probability distribution is
\begin{equation}
    g(z) = \begin{cases}   
    \frac{1}{\sqrt{z}}, & \text{if $z \in [1/a,a]$}\\
    0, & \text{otherwise}, 
    \end{cases}
    \label{eq:g_stretch}
\end{equation}
with $a$ arbitrarily set to 2~\cite{goodmanweare}.
The new proposed point is then accepted with probability
\begin{equation}
    \alpha (X_{i, {\rm new}}, X_i(t)) = \min \left\{1,  z^{d-1} \frac{p(X_{i, {\rm new}})}{p(X_i(t))}\right\},
    \label{eq:acc_stretch}
\end{equation}
where $p(X)$ denotes the target probability density, evaluated at the point $X$, $d$ is the dimension of the parameter space involved in the stretch move, and $z^{d-1}$ a factor needed to preserve the symmetry condition; see Appendix~\ref{ssec:app_symm_stretch} for details. 

When choosing the other random walker $X_j$ needed for the stretch move, one must ensure that the position of this walker is not updated at the same time as the move we are currently attempting. 
In the group stretch move in \texttt{eryn}, the walkers are split into two (or more) separate groups and are updated sequentially. 
When updating the positions of walkers in a given group, the random walkers $X_j$ are then chosen from the other one(s), therefore called \textit{complementary}. 

As explained in detail later, we will employ stretch moves in \texttt{t-roo} for different purposes: to update the single models' parameters and for the proposal of tidal deformability parameters when considering jumps between models (cf. Sec.~\ref{sec:proposals}).

\subsection{Parallel tempering}
\label{sec:temp}
MCMC samplers can struggle with complex and multimodal distributions since there is a possibility that walkers get stuck in one of the modes. 
One way to mitigate this issue is through \emph{parallel tempering}~\cite{parallel_tempering,01d8291b-8e72-3d7a-9509-e149ce24afee,2005PCCP....7.3910E}. 
In parallel tempering, for each walker $n_T$ chains are run in parallel, each one with a different temperature $T_j$, which transforms the likelihood as
\begin{equation}
    \mathcal{L} (d| \vec{\theta},M) \rightarrow \mathcal{L}(d|\vec{\theta},M) ^{1/T_j},
\end{equation}
meaning that the $j$-th chain is effectively exploring the posterior
\begin{equation}
    p_{T_j}(\vec{\theta}, d) \propto  \mathcal{L}(d|\vec{\theta},M) ^{1/T_j} \pi (\vec{\theta}).
\end{equation}
Basically, the temperature effect ``smoothens" the likelihood profile, allowing the walkers to explore low-likelihood regions of the parameter space more easily. 
The chain with $T_j = 1$ explores the ``true" posterior distribution, i.e., the target distribution, while the $T_j = \infty$ chain explores the prior. Different chains with different temperatures run in parallel and periodically exchange information. 
Chains with higher temperatures sweep across the parameter space, discovering (new) regions with potentially high likelihood, and then pass this information to the low-temperature chains, which can better explore the likelihood peaks. 
In order to preserve detailed balance, the swaps between chains with different temperatures are accepted with a probability
\begin{equation}
    \alpha(T_i, T_j) = \min \left\{ 1, \left( \frac{\mathcal{L}(d|\vec{\theta}_i)}{\mathcal{L} (d|\vec{\theta}_j) } \right)^{\frac{1}{T_j} - \frac{1}{T_i}} \right\}.
\end{equation}

The number of required temperature chains depends on the problem at hand, but usually at least 10-20 are needed. 
The choice of the specific temperature values is non-trivial, since ideally it should maximize the information exchange among chains. 
To find the optimal temperature ladder, \texttt{eryn}, and consequently \texttt{t-roo}, employ the dynamic adaptation algorithm developed in Ref.~\cite{2016MNRAS.455.1919V}.
It is good practice to consider the samples accumulated during the temperature optimization as part of the burn-in. 
In \texttt{t-roo}, temperature adaptation happens only during the preliminary sampling stage (see Sec.~\ref{sec:mass_proposal}) and is automatically switched off during the RJMCMC phase. Therefore, when extracting the odds ratio or posterior parameters from the RJMCMC iterations, these samples are automatically not taken into account.
As we will see later in Appendix~\ref{sec:evidences}, parallel tempering also allows to compute evidences in MCMC samplers.

\begin{figure*}[tbh]
        \centering        
        \includegraphics[width=1\textwidth]{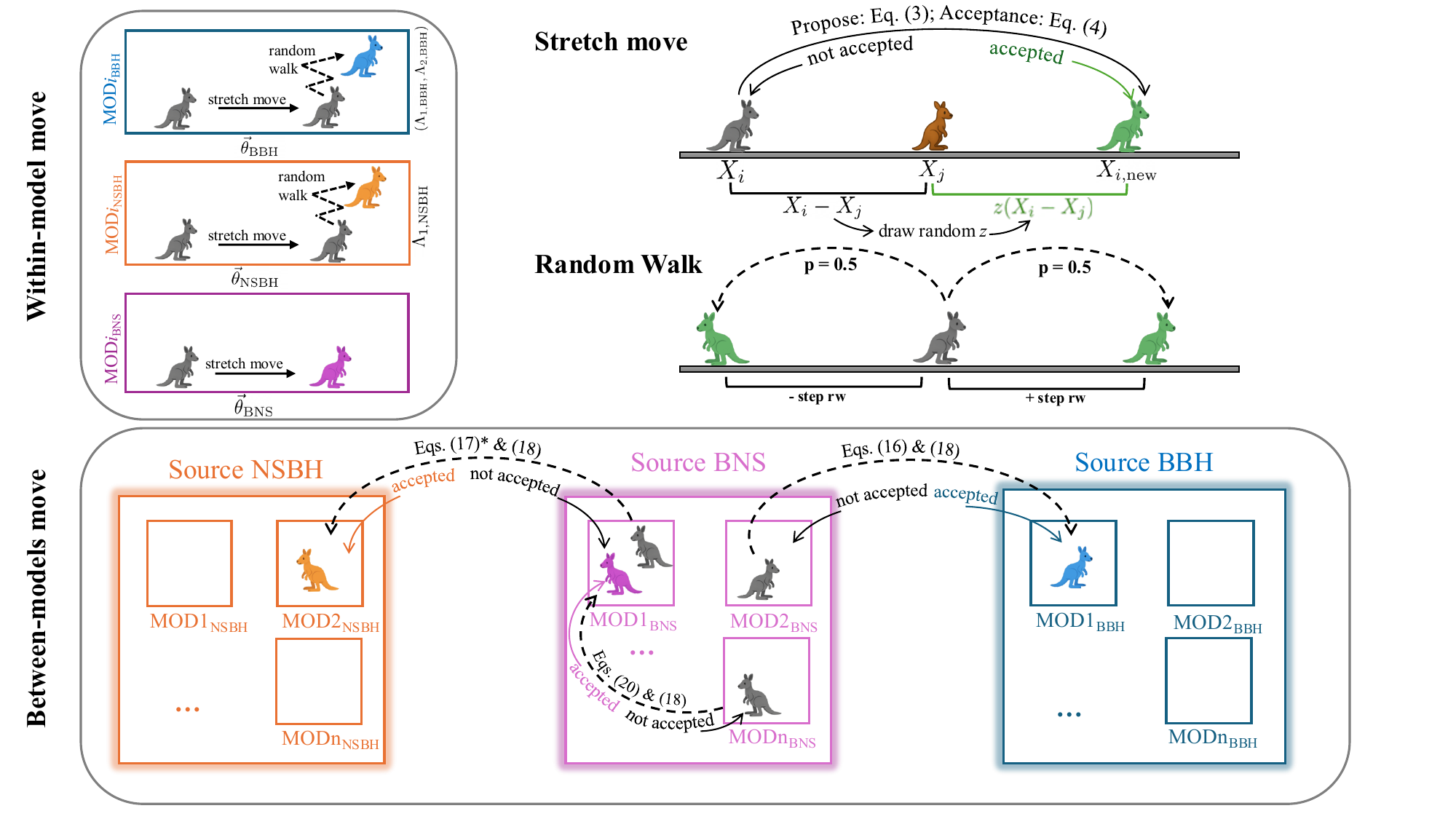}
    \caption{Schematic representation of the moves implemented in \texttt{t-roo}, with reference to the respective equations in the text; the asterisk $^*$ denotes the inverse transformation. For the between-model move, the walkers can jump to any model, of a different or the same source class, with the same probability. For the within-model move, we update the \textit{pseudo}-parameters with a random walk with step ``step rw'' and all the other parameters with a stretch move.
    See Sec.~\ref{sec:proposals} for a complete discussion of the proposals.}
    \label{fig:troo_rjmcmc}
\end{figure*}

\subsection{Bilby features}
As mentioned in the introduction, \texttt{t-roo} employs the sampler infrastructure of \texttt{eryn}, combining it with \textsc{Bilby} features in order to analyze CBC signals. 
In particular, the detector classes, the waveform generators, and the likelihood are directly called from \textsc{Bilby} with specific wrappers, which allow us to import advanced features implemented there, e.g., likelihood marginalization~\cite{marg}, or injections in different synthetic noise realizations.

For the runs presented in this work, we employ the \emph{multibanding} approximation of the likelihood~\cite{Morisaki:2021ngj}, which restricts the waveform evaluation to a reduced set of frequency points, adapted according to the signal duration, hence significantly reducing the computational cost. However, we note that in principle all the likelihood methods --- e.g., the ``standard" \texttt{GravitationalWaveTransient} likelihood, the relative binning~\cite{Zackay:2018qdy}, or the reduced-order-quadrature (ROQ)~\cite{Smith:2016qas} ones--- can be employed in \texttt{t-roo}\footnote{In the case of ROQ, this would would require the ROQ data to be available in the needed mass range for all the models investigated.}. 
Similarly, the prior infrastructure is heavily based on \textsc{bilby}, therefore also somewhat specific priors such as the aligned spin one for the spin magnitudes or the uniform in comoving volume for the luminosity distance are available in \texttt{t-roo}.

\section{t-roo proposals}
\label{sec:proposals}

As discussed in detail in Appendix~\ref{sec:rjmcmc_theory}, an RJMCMC sampler performs two different moves during each sampling step\footnote{One could also decide to have multiple within-model moves in a sampling step. This possibility is implemented in \texttt{t-roo} but not used by default, since the preliminary sampling already helps identify the parameter regions with the largest support for each model.}. For each walker we have:

\begin{enumerate}
    \item \emph{Within-model move:} propose new parameter values for the model the walker is currently in and decide whether to accept the move to this new point in this parameter space or not.
    \item \emph{Between-models move:} propose a jump to a different model, and simultaneously update the parameters in the new model; also in this case, there is a criterion to decide whether to accept the proposed jump or not.
\end{enumerate}

In the following, we describe the specific moves that we implemented in \texttt{t-roo}.
We highlight that one strength of our approach is that we are not only able to compare different kinds of compact binaries, but we can also simultaneously rank multiple waveform models within a certain binary category. 
Moreover, we can directly jump between BNS and BBH models, without the need to go through a NSBH one, which reduces the number of iterations and, consequently, the computational cost.
Figure~\ref{fig:troo_rjmcmc} shows a schematic overview of the \texttt{t-roo} algorithm and the different moves involved.

The proposals are written assuming sampling in chirp mass $\mathcal{M}_c$, mass ratio $q=m_2/m_1$, and component tidal deformabilities $\Lambda_1$ and $\Lambda_2$. 
Although, in principle, it would be possible to sample over different combinations of these parameters and convert them before the proposal, this would yield an (unnecessary) computational cost. 
The posterior probabilities corresponding to different choices of sampling parameters can anyway be obtained by reweighting the final posteriors.

\subsection{Between-models move}
\label{subsec:between_models_move}

The choice for a specific form of the between-model moves is the hardest challenge in RJMCMC algorithms, because their acceptance ratio strongly affects the sampler's overall performance.   

The proposed values for the parameters in the new model depend on the parameters in the current one: to avoid the risk of updating the two states simultaneously, in between-model jumps, we update all walkers in a given model at the time, proceeding sequentially through the models. This approach is similar to the idea of a \textit{group} stretch move, since the walkers are effectively divided into different groups that are updated in sequence. The difference is that the groups are not created with a random partition, but based on the model the walkers are currently in.

When more than two models are compared, \texttt{t-roo} randomly chooses which model a jump is proposed to, assuming that all the competing models have an equal probability.  
One could take into account prior information about the preferred models or source type by assigning models priors, that could be alternatively assigned (or further divided) based on the source classification.
This would enable also the incorporation of astrophysical information, e.g., what is the expected probability to detect systems with specific features, such as eccentricity or precession.
This is left for future work, and currently the models' priors can be incorporated a-posteriori by multiplying the final odds ratios.

\subsubsection{Tidal deformability}

Since our primary goal is to compare models that describe BBH, BNS, and NSBH systems, the core of the between-model proposal concerns the tidal deformability parameters, $\Lambda_1$ and $\Lambda_2$. 
In principle, one could simply add and remove $\Lambda_1$ and $\Lambda_2$ based on the model jump (for example, when going from NSBH to BBH we could just remove $\Lambda_2$, while when going from BBH to BNS we could add $\Lambda_1$ and $\Lambda_2$). 
However, this would mean losing the tidal information gained while sampling with one model when going to a model with fewer tidal parameters. 
To avoid this, we add the \textit{pseudo} tidal parameters \lonebb, \ltwobb, and \lonenb{} for BBH and NSBH models, which essentially act as ``storage" variables. 
Crucially, these \textit{pseudo}-parameters are not included in the likelihood computation for the model they belong to, but serve just to cache the tidal information until the next jump to a model where that tidal parameter is actually included.

The proposal between tidal parameters in different models is a \emph{stretch} move, the primary move employed in \texttt{emcee} and explained in Sec.~\ref{sec:stretch_move}.
In the implementation proposed by Goodman and Weare in Ref.~\cite{goodmanweare}, the random variable $z$ defining the stretch move is sampled in $[1/a,a]$, with $a$ arbitrarily set to $a=2$.
However, in our between-model moves, the \textit{pseudo}-parameters are used to store the tidal information, thus, ideally, we do not want this stretch move to propose too large jumps. Therefore, we choose
\begin{equation}
    g(z) = \begin{cases}   
    \frac{1}{\sqrt{z}}, & \text{if $z \in [1-\epsilon, 1+\epsilon]$},\\
    0, & \text{otherwise}, 
    \end{cases}
    \label{eq:udistr}
\end{equation}
with $\epsilon$ set by default to $0.01$, but adjustable by the user. 
Proposing very small jumps, of the order of a few percent, effectively means that we collect most of the information about the tidal parameters during the within-model moves in NSBH or BNS models. $X_j$ is picked from the walkers in the other models, i.e., the ones that are currently not updated. In order to avoid unphysical values of tidal deformability, if the proposed value becomes negative, we simply retain the old one. 

\subsubsection{Proposals for mass parameters}
\label{sec:mass_proposal}
\begin{figure*}[htb]
        \centering
        \includegraphics[width=1.\linewidth]{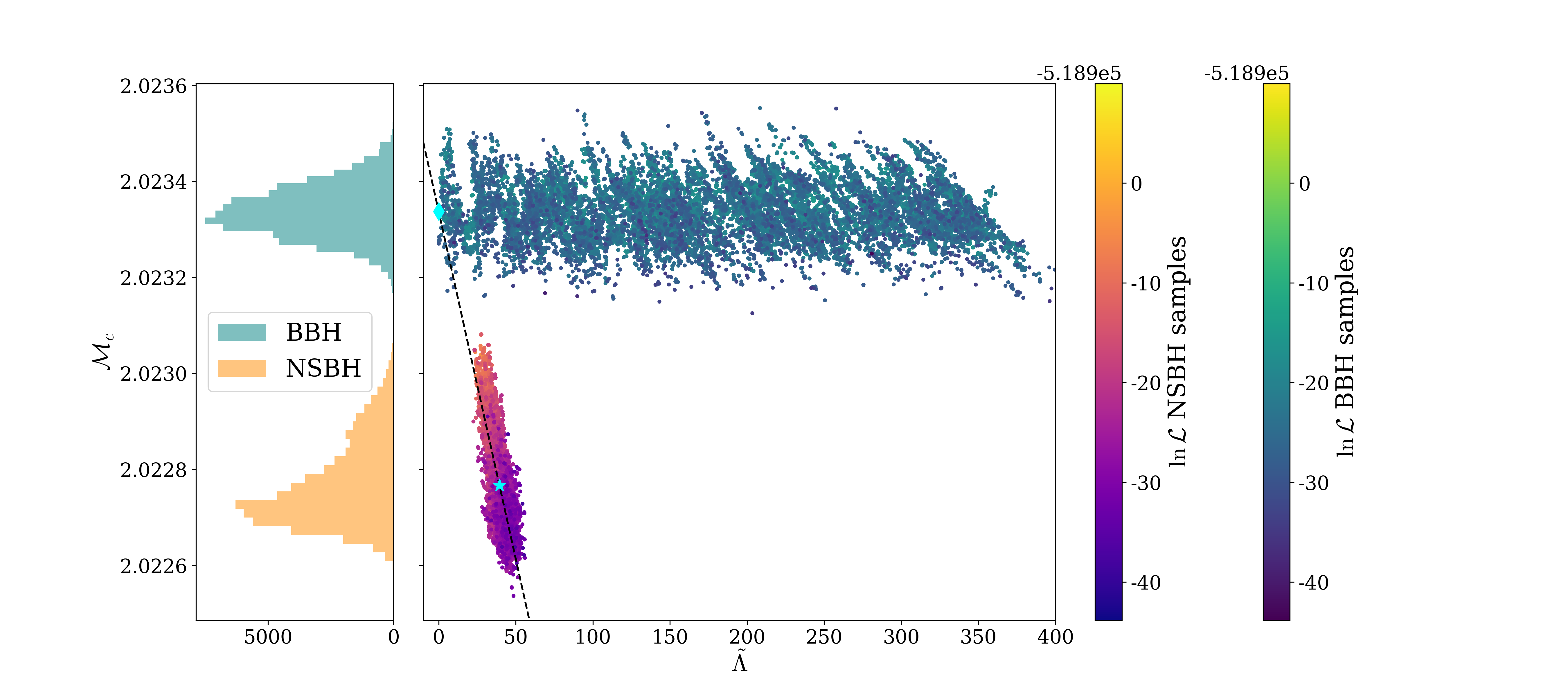}
    \caption{Example of $\mathcal{M}_c$ and $\tilde{\Lambda}$ distribution of samples collected by all the walkers in the last 200 steps of the preliminary sampling phase in the case of a NSBH injection with $\Lambda_2 =300$ as recovered by ET (\textsc{nsbh\_lam300\_et}, see Sec.~\ref{ssec:inj} for details), color-coded by the likelihood value of each sample; the orange-red colors correspond to samples recovered with the NSBH approximant \texttt{IMRPhenomNSBH} and the blue-green with the BBH approximant \texttt{IMRPhenomD}. The left panel displays the one-dimensional $\mathcal{M}_c$ distribution for the BBH and NSBH case, showing how the recovered support lies in disjoint ranges for the two models.
    We note that in the BBH model \lonebb{} and \ltwobb, from which the $\tilde{\Lambda}$ parameter shown here is computed, are \textit{pseudo-}parameters used merely as storage variables, and therefore do not affect the likelihood.  
    The cyan diamond and star indicate the likelihood center-of-mass values for the BBH and NSBH model, respectively. The proposal slope in Eq.~\eqref{eq:slope_mc} is computed from the dashed line connecting them.} 
    \label{fig:mc_proposal}
\end{figure*}
The presence of tidal effects in a signal can alter which values for $\mathcal{M}_c$ and $q$ maximize the likelihood, because of the accelerated inspiral (due to the additional dissipated energy) and the degeneracy between tidal parameters and mass ratio. 
This circumstance might cause the sampler to get stuck in one of the models: if we propose the jump to a different model assuming the same values for the mass parameters, the new model might not have significant likelihood support at those masses, especially for high-SNR signals.
Therefore, we implement specific data-informed proposals for chirp mass and mass-ratio, which take into account the effect of different tidal contents\footnote{We only look empirically at the effect of including or not tidal phase contributions in the model, but a more complete proposal would need to take into account other contributions of matter effects, for example the differences in the spin-spin term in the phase description due to the different quadrupole moment values in NSs and BHs.}.
 
A \texttt{t-roo} run starts with a \textit{preliminary sampling} phase, during which the sampler runs on each model separately for some time without proposing between-model moves\footnote{Note that parallel tempering is enabled during this stage, therefore swaps between models due to tempering are still possible.}.

Figure~\ref{fig:mc_proposal} shows an example of the distribution of $\mathcal{M}_c$ and $\tilde{\Lambda}$ samples after the preliminary sampling, illustrating how the NSBH and BBH model find disjoint chirp mass intervals.
To identify a mapping between these different intervals, and similarly for the mass ratio, the sampler considers the $\mathcal{M}_c$, $q$, and $\tilde{\Lambda}$ samples from each model after the preliminary sampling phase has finished, with $\tilde{\Lambda}$ being the mass-weighted (or ``reduced") tidal deformability parameter~\cite{Favata:2013rwa}. 
It determines the likelihood ``center-of-mass`` of the $\mathcal{M}_c$, $q$, and $\tilde{\Lambda}$ samples collected during the last 200 steps:
\begin{align}
   \mathcal{M}_{c, \rm \textsc{cm}} &= \frac{\sum_i \mathcal{L}(\vec{\theta}_i) \mathcal{M}_{c,i}}{\sum_i \mathcal{L}(\vec{\theta_i})}, \\
   q_{\rm \textsc{cm}} &= \frac{\sum_i \mathcal{L}(\vec{\theta}_i) q_{i}}{\sum_i \mathcal{L}(\vec{\theta_i})}, \\
   \bar{\tilde{\Lambda}}_{\rm \textsc{cm}} &= \frac{\sum_i \mathcal{L}(\vec{\theta}_i) \tilde{\Lambda}_{i}}{\sum_i \mathcal{L}(\vec{\theta_i})},
\end{align}
Here, the index $i$ runs over the model's samples, and $\vec{\theta}$ refers to the parameters of that specific model. 
Furthermore, the bar on $\tilde{\Lambda}$ denotes that for the computation of $\tilde{\Lambda}_{\rm \textsc{cm}}$ in the BBH and NSBH models, we assume the physical value of zero tidal deformability for the BH object, i.e., \lonebb=\ltwobb=\ltwonb=0 (yielding, for BBH models, $\tilde{\Lambda}_{\textsc{cm}}=0$), because these are the values employed in the likelihood calculation.
We consider only the last 200 iterations to ensure no samples from the burn-in phase are used. 
However, how many final iterations should be used can be adjusted by the user.
Using the equivalent of the center of the mass of the likelihood instead of, for example, the sample corresponding to its maximum value, ensures that we take into account the overall trend of the distribution, making the point choice more robust against potential biases or outliers.

We consider the line connecting the different likelihood-center-of-mass points in the $\mathcal{M}_c$-$q$-$\tilde{\Lambda}$ space.
For two different models $\textsc{mod}_1$ and $\textsc{mod}_2$, its slope is given by 
\begin{align}
    & s_{\mathcal{M}_c (\textsc{mod}_1, \textsc{mod}_2)} = \frac{\mathcal{M}_{c, \textsc{cm}, \textsc{mod}1} - \mathcal{M}_{c, \textsc{cm}, \textsc{mod}_2}}{\bar{\tilde{\Lambda}}_{\textsc{cm},\textsc{mod}_1} - {\bar{\tilde{\Lambda}}_{\textsc{cm},\textsc{mod}_2}}}  \label{eq:slope_mc}\\
    & s_{q (\textsc{mod}_1, \textsc{mod}_2)} = \frac{q_{ \textsc{cm}, \textsc{mod}_1} - q_{\textsc{cm}, \textsc{mod}_2}}{\bar{\tilde{\Lambda}}_{\textsc{cm},\textsc{mod}_1} - {\bar{\tilde{\Lambda}}_{\textsc{cm}, \textsc{mod}_2}}}. \label{eq:slope_q}
\end{align}
For each between-models move, the proposed $\mathcal{M}_c$ and $q$ for the new model are computed following the lines with these slopes, starting from the point defined by the current sample in the current model, and assuming the physical values \lonebb$=0$, \ltwobb$=0$, and \lonenb$=0$.

However, if the ranges for the $\mathcal{M}_c$ support in different models overlap sufficiently, this specific proposal is not needed.
In this case, the chirp mass value is not updated in the between-model move\footnote{We skip the chirp mass proposal mostly to save computational time, although we note that the main computational bottleneck is in any case the waveform evaluation.}. 
Specifically, when going from $\textsc{mod}_1$ to $\textsc{mod}_2$, the chirp-mass proposal is \textit{not} employed when the likelihood center-of-mass value of $\mathcal{M}_c$ for $\textsc{mod}_2$ lies within the $68\%$ interval of the $\mathcal{M}_c$ samples from $\textsc{mod}_1$.
We do not apply this condition to the mass ratio, because the posteriors for this parameter are usually much wider and this distinguishability criterion will rarely be satisfied.

Besides identifying the right map between the different model parameters, the preliminary sampling phase also improves the overall convergence of the sampler.
Before the reversible jump even starts, the walkers will already have reduced the region of parameter space to explore for each model. 
The number of iterations in the preliminary sampling phase should be enough to pass the burn-in phase and provide reliable, although not definitive, posteriors for the parameters of interest, but not too long to affect the computational performance of the algorithm.

\subsubsection{The complete proposal}

Taken all together, the between-model proposals read (see Appendix~\ref{sec:app_proposals} for the derivation)

\begin{align}
& \qquad \textbf{BBH to NSBH} \nonumber\\
&\begin{cases}
        \vec{\Theta}_{\textsc{nsbh}, i+1} = \vec{\Theta}_{\textsc{bbh}, i} \\
        \Lambda_{1, \textsc{nsbh}, i+1} = \Lambda_{1, \textsc{bbh}, i} \\
        \Lambda_{2, \textsc{nsbh}, i+1} = \Lambda_{2,j, i} + u_2 \left( \Lambda_{2, \textsc{bbh}, i} - \Lambda_{2,j, i} \right) \\
        \mathcal{M}_{c, \textsc{nsbh}, i+1} = \mathcal{M}_{c, \textsc{bbh}, i} + s_{\mathcal{M}_c (\textsc{nsbh}, \textsc{bbh})} \cdot \bar{\tilde{\Lambda}}_{\textsc{nsbh}, i+1} \\
        q_{\textsc{nsbh}, i+1} = q_{\textsc{bbh}, i} + s_{q (\textsc{nsbh}, \textsc{bbh})} \cdot \bar{\tilde{\Lambda}}_{\textsc{nsbh}, i+1} \\
        v_2 = 1/u_2 
        \end{cases}  \label{eq:bbh_to_nsbh} \\[3ex]
&\qquad \textbf{BNS to BBH} \nonumber\\
&    \begin{cases}
        \vec{\Theta}_{\textsc{bbh}, i+1} = \vec{\Theta}_{\textsc{bns}, i} \\
        \Lambda_{1, \textsc{bbh}, i+1} = \Lambda_{1,j, i} + w_{1} \left( \Lambda_{1, \textsc{bns}, i} - \Lambda_{1,j, i} \right) \\
        \Lambda_{2, \textsc{bbh}, i+1} = \Lambda_{2,j, i} + w_2 \left( \Lambda_{1, \textsc{bns}, i} - \Lambda_{2,j, i} \right) \\
        \mathcal{M}_{c, \textsc{bbh}, i+1} = \mathcal{M}_{c, \textsc{bns}, i} - s_{\mathcal{M}_c (\textsc{bbh}, \textsc{bns})} \tilde{\Lambda}_{\textsc{bns}, i} \\
        q_{\textsc{bbh}, i+1} = q_{\textsc{bns}, i} - s_{q (\textsc{bbh, \textsc{bns})}} \tilde{\Lambda}_{\textsc{bns}, i} \\
        u_1 = 1 / w_1 \\
        u_2 = 1/w_2 
        \end{cases} \label{eq:bns_to_bbh}
\end{align}

\begin{align}
&\qquad \textbf{NSBH to BNS} \nonumber\\
&\begin{cases}
        \vec{\Theta}_{\textsc{bns}, i+1} = \vec{\Theta}_{\textsc{nsbh},i} \\
        \Lambda_{1, \textsc{bns}, i+1} = \Lambda_{1,j,i} + v_1 \left( \Lambda_{1, \textsc{nsbh}, i} - \Lambda_{1,j, i} \right) \\
        \Lambda_{2, \textsc{bns}, i+1} = \Lambda_{2,j,i} + v_2 \left( \Lambda_{1, \textsc{nsbh}, i} - \Lambda_{2,j, i} \right) \\
        \begin{aligned}
        \mathcal{M}_{c, \textsc{bns}, i+1} = &\mathcal{M}_{c, \textsc{nsbh}, i} + \\ &- s_{\mathcal{M}_c (\textsc{bns}, \textsc{nsbh})}  \left( \bar{\tilde{\Lambda}}_{\textsc{nsbh}, i} - {\tilde{\Lambda}}_{\textsc{bns}, i+1} \right) \end{aligned} \\
        q_{\textsc{bns}, i+1} = q_{\textsc{nsbh}, i} - s_{q (\textsc{bns}, \textsc{nsbh})}  \left( \bar{\tilde{\Lambda}}_{\textsc{nsbh}, i} - {\tilde{\Lambda}}_{\textsc{bns}, i+1} \right) \\
        w_1 = 1 / v_1 \\
        w_2 = 1/v_2
        \end{cases} \label{eq:nsbh_to_bns}
\end{align}
for the $\textsc{bbh} \rightarrow \textsc{nsbh}$, $\textsc{bns} \rightarrow \textsc{bbh}$, and $\textsc{nsbh} \rightarrow \textsc{bns}$ moves, respectively. 
The subscript $i$ denotes current samples, while $i+1$ the proposed ones,  and
$j$ refers to the complementary set of samples used for the stretch move.
The parameter vector $\vec{\Theta}$ stands for the parameters for which no specific proposal is implemented, i.e., $\vec{\Theta} = \vec{\theta} \; \setminus \; \left\{ \mathcal{M}_c, q, \Lambda_1, \Lambda_2 \right\} $.
In the chirp-mass and mass-ratio proposal, the barred $\bar{\tilde{\Lambda}}$ indicate that, in the case of NSBH or BBH systems, we do not use the stored values of the \textit{pseudo}-parameters for the computation of $\tilde{\Lambda}$, but we assume $\Lambda_{1, \textsc{bbh}} = \Lambda_{2, \textsc{bbh}} = \Lambda_{1, \textsc{nsbh}} = 0$, i.e., we assume physical values of zero tidal deformability for BHs, as for the computation of the slopes in Eqs.~\eqref{eq:slope_mc}-\eqref{eq:slope_q}.
The proposal for the auxiliary variables $u_1, u_2, v_1, v_2, w_1, w_2$ is determined by the requirement that the transformations must be reversible (see Appendix~\ref{ssec:tdim_mcmc} and \ref{sec:app_proposals} for details). 
The auxiliary variables $w_1$ and $v_1$ are not needed when swapping between NSBH and BBH models, because \lonebb{} and \lonenb{} are both \textit{pseudo}-parameters that can simply be set equal to each other, skipping the stretch-move.
The opposite transformations, i.e., $\textsc{nsbh} \rightarrow \textsc{bbh}$, $\textsc{bbh} \rightarrow \textsc{bns}$, and $\textsc{bns} \rightarrow \textsc{nsbh}$, are simply the inverse ones, consistent with the requirements that, in RJMCMC algorithms, each proposal should be reversible (the inverse mapping is provided in Appendix~\ref{sec:app_proposals}).

Once the new model and parameters values have been proposed, they are accepted according to the acceptance probability 
\begin{equation}
\alpha(x, x') = \min \left\{ 1, \frac{\mathcal{L}\left( \vec{\theta}_{x'}\right) p \left( \vec{\theta}_{x'}\right) g'_{x'}} {\mathcal{L}\left( \vec{\theta}_x\right) p \left( \vec{\theta}_x \right) g_x } \mathbb{J}_{xx'}  \right\},
\label{eq:between_acc}
\end{equation}
where $x$ denotes the ``old" state model and parameters at step $i$, while $x'$ the new model and parameters of the proposed state at the following step $i+1$. $g_x$ and $g_{x'}$ represent the probability distributions of the auxiliary variables for the model in the states $x$ and $x'$, respectively (see Appendix~\ref{sec:rjmcmc_theory} for details). $\mathcal{L}(\vec{\theta}_y)$ and $p(\vec{\theta}_y)$ denote the likelihood and prior for the parameters and model in stare $y$. 
As explained in detail in Appendix~\ref{ssec:app_priors}, in order to preserve the detailed balance, the priors $p (\vec{\theta}_x )$, or $p ( \vec{\theta}_{x'} )$, must incorporate all the parameters we are sampling over, including the \textit{pseudo-}parameters \lonebb, \lonenb, and \ltwobb, even if they are not used in the likelihood calculation.

For the proposals in~\Cref{eq:bbh_to_nsbh,eq:bns_to_bbh,eq:nsbh_to_bns} and their inverses, the Jacobians $\mathbb{J}_{xx'}$ 
depend exclusively on the auxiliary variables, and, thanks to the specific mapping between auxiliary variables in the different models,
\begin{equation}
  \frac{g_{x'}}{g_x}  \mathbb{J}_{xx'} = 1
  \label{eq:factor}
\end{equation}
in all cases; see Appendix~\ref{app:jacobians} for the computation.

\subsubsection{Proposal between models for the same type of source and parameter space}

\texttt{t-roo} provides the possibility to compare not only models that describe different kinds of binary systems, but also multiple models for a given source type. In this case, we do not need to take into account the effects of different tidal contents, therefore the parameter transformation between one model and the other is simply an identity transformation

\begin{equation}
    \vec{\theta}_{\textsc{mod}_2} = \mathbf{I} \vec{\theta}_{\textsc{mod}_1},
\end{equation}
with, consequently, a Jacobian $\mathbb{J}_{xx'} = 1$.

We note that this is possible because, for the time being, we are assuming only models with the same physics, for example, aligned-spin only. Comparing models describing not only different sources, but also different features for the same source, such as the presence of precession, or eccentricity, would require dedicated proposals for the between-model moves, and is left for future work.

\subsection{Within-model moves}

We update the parameters within a model with a group stretch move as implemented in \texttt{eryn} and described in Sec.~\ref{sec:stretch_move}, with the exception of the \textit{pseudo}-parameters \lonebb, \lonenb, and \ltwobb{}.
Given that they are included in the between-model moves, the \textit{pseudo-}parameters are effectively part of the sampled parameter spaces, and as such have to be updated also in the within-model moves. However, since they are not relevant for the likelihood computation, we do not expect to gain information about them. Thus their posterior should ideally recover the prior, which, for simplicity, we assume to be the same as the actual tidal parameters in the BNS and NSBH models, i.e., uniform in the component tidal deformabilities.
To facilitate the recovery of uniform distributions, we update the \textit{pseudo}-parameters in the within-model move through a random walk with a default step size of 10 (in units of dimensionless tidal deformability).
If the walkers spend a sufficient amount of time in a model without tidal information, the random walk ensures that the posterior samples of the \textit{pseudo}-parameters will be uniformly distributed in the prior range.

We include the possibility of \emph{block-sampling} in the within-model moves, meaning that we can update only specific subsets of the parameters at each step.
We employ the same structure as in the \textsc{Bilby}-MCMC proposal cycle~\cite{Ashton:2021anp}, but in our case only the parameters involved in the proposal change, while the move remains the same, i.e., a stretch move, or a random walk for \lonebb, \ltwobb, or \lonenb{}.
Block sampling is particularly useful in case of multi-modal distributions, since it allows the walkers to better explore the full range of each parameter. 
However, one of the strengths of affine samplers, such as \texttt{emcee}, on which \texttt{eryn}, and consequently \texttt{t-roo}, are based, is their efficiency in dealing with highly-correlated parameters, which is the case for CBC GW signals. 
Therefore, in this case, employing block sampling can actually slow-down convergence, and it is by default not used in \texttt{t-roo}.

\section{Analysis methods}
\label{sec:methods}

\begin{table*}[!ht]
\begingroup
\caption{Injections analyzed to assess the sampler's performance. Together with the labels, we provide the component masses, spins, and tidal deformabilities. 
The second last column reports the models used to create the mock signal for the injection, while the last one lists the models employed for recovery.}

\label{tab:injections}
\renewcommand*{\arraystretch}{2}
\setlength{\tabcolsep}{6pt}
\begin{tabular}{l c c c c c}
\hline
        & $m_1$, $m_2$ [$M_\odot$] & $\chi_1$, $\chi_2$ & $\Lambda_1$, $\Lambda_2$ & Injected model & Recovery models \\
        \hline
    \textsc{nsbh\_lam300} & 3.75, 1.5 & 0.6, 0.02 & 0, 300 & \texttt{IMRPhenomNSBH} & \makecell[t]{\texttt{IMRPhenomNSBH} \\ \texttt{IMRPhenomD}} \\
    \textsc{nsbh\_lam600} & 3.75, 1.5 & 0.6, 0.02 & 0, 600 & \texttt{IMRPhenomNSBH} & \makecell[t]{\texttt{IMRPhenomNSBH} \\ \texttt{IMRPhenomD}} \\
    \textsc{nsbh\_lam1000} & 3.75, 1.5 & 0.6, 0.02 & 0, 1000  & \texttt{IMRPhenomNSBH} & \makecell[t]{\texttt{IMRPhenomNSBH} \\ \texttt{IMRPhenomD}} \\
    \textsc{bbh\_nsbhmasses} & 3.75, 1.5 & 0.6, 0.02 & 0, 0 &  \texttt{IMRPhenomD} & \makecell[t]{\texttt{IMRPhenomNSBH} \\ \texttt{IMRPhenomD}} \\
    \textsc{bns\_lam600} & 1.4, 1.4 & 0.01, 0.02 & 600, 600 &  \texttt{IMRPhenomD\_NRTidalv2} & \makecell[t]{\texttt{IMRPhenomD\_NRTidalv2} \\ \texttt{IMRPhenomNSBH} \\ \texttt{IMRPhenomD}} \\
    \textsc{bbh\_bnsmasses} & 1.4, 1.4 & 0.01, 0.02 & 0, 0 &  \texttt{IMRPhenomD} & \makecell[t]{\texttt{IMRPhenomD\_NRTidalv2} \\ \texttt{IMRPhenomNSBH} \\ \texttt{IMRPhenomD}} \\
    \textsc{bns\_lam600\_mm} & 1.4, 1.4 & 0.01, 0.02 & 600, 600 &  \texttt{IMRPhenomXAS\_NRTidalv3} & \makecell[t]{\texttt{IMRPhenomD\_NRTidalv2} \\ \texttt{IMRPhenomXAS\_NRTidalv3} \\ \texttt{SEOBNRv5\_ROM\_NRTidalv3} \\ \texttt{IMRPhenomD} \\ \texttt{IMRPhenomXAS} \\ \texttt{SEOBNRv5\_ROM}} \\

\end{tabular}
\endgroup
\end{table*}

\subsection{Injections setup}
\label{ssec:inj}

To assess the robustness of the sampler, we test it on simulated data obtained by injecting different mock GW signals on top of synthetic detector noise. 
We assume a detector network composed of the two Advanced LIGO detectors, Hanford and Livingston~\cite{LIGOScientific:2014pky}, and the Advanced Virgo detector~\cite{VIRGO:2014yos} at design sensitivity~\cite{LIGO-T2000012v1, LIGO2020ObservingScenarios}. 
We consider NSBH signals with different tidal contributions and recover them with an NSBH and a BBH model, to investigate when \texttt{t-roo} can identify the presence of a NS component. 
We also analyze BNS injections and recover them using different models describing BNS, NSBH, and BBH systems. 
In addition, we look at signals from BBH sources, with masses compatible with either BNS or NSBH systems, to evaluate how the sampler behaves in the absence of tidal information. 
All injections are summarized with their labels and source properties in Table~\ref{tab:injections}.

For NSBH sources, we use the \texttt{IMRPhenomNSBH} approximant~\cite{Thompson:2020nei}, which is based on the BBH amplitude prescription in Ref.~\cite{Pannarale:2015jka}, in turn based on \texttt{IMRPhenomC}, and the tidal phase description in \texttt{IMRPhenomD}~\cite{Khan:2015jqa} and \texttt{NRTidalv2}~\cite{Dietrich:2019kaq}. 
Therefore, the comparison is performed with the \texttt{IMRPhenomD} approximant for the BBH case and \texttt{IMRPhenomD\_NRTidalv2} for the BNS one, in order to reduce potential modelling systematics.
On the other hand, when comparing multiple models for each source, we consider also other state-of-the-art approximants~\cite{Pratten:2020fqn, Pompili:2023tna, Abac:2023ujg}.

For all injections, we place the source at a luminosity distance of $d_L = 30$~Mpc in order to have a large signal-to-noise ratio, which enables us to identify tidal signatures if present. 
SNRs range roughly between 50 and 100 depending on the simulated system. 
For the injection \textsc{nsbh\_lam600}, we perform an additional run at $d_L = 150~$Mpc (\textsc{nsbh\_lam600\_lowsnr}), to test the performance at lower SNR (specifically, this signal has an SNR of 14).
We also perform the \textsc{nsbh\_lam300} and \textsc{bns\_lam600} injections for an ET setup (\textsc{nsbh\_lam300\_et} and \textsc{bns\_lam600\_et}, respectively), assuming two L-shaped detectors with a 15~km arm-length and misaligned orientation, placed in the proposed Sardinia and Lusatia sites~\cite{Branchesi:2023mws, ET:2025xjr, et_psd}.

We employ the multibanding likelihood, as implemented in \textsc{Bilby}, together with phase and distance marginalization in order to reduce the computational cost.
Signals are analyzed starting from 20~Hz, and in all the tests we use $N_w = 500$ walkers, each with $16$ temperature chains.
These are implemented as the default values in \texttt{t-roo}, but can be changed depending on the analysis requirements. 
In particular, when sampling with a large number of models, it is advisable to increase the number of walkers, to ensure an adequate exploration of all parameter spaces. 

\subsection{Priors}

The parameter priors are kept the same across different models, in order to avoid penalizing or favoring models with different prior support. 
For chirp mass, we choose a uniform prior in $[1.15, 1.30] \, M_\odot$ and $[1.95,2.10] \, M_\odot$ for the injections of sources with masses compatible with BNS systems or not, respectively. 
In the analysis of real events, chirp mass is sampled uniformly in the range $[1.45,1.52] \, M_\odot $ and $[2.02,2.04] \, M_\odot$ in the case of GW190425 and GW230529, respectively.
We use BNS models in the recovery only if the injected masses are fully compatible with BNS systems, meaning that we demand the full prior range to lie in the region of validity of the model\footnote{In the case of real events, one would refer to low-latency point estimates, or to preliminary parameter estimation results.}.
For signals with masses close to the maximum value allowed for neutron star systems, the wider mass prior would cause the waveform evaluation for the BNS models to fail, thus artificially disfavoring the model. 
Although intuitively correct, since for quite large masses we expect the BNS hypothesis to be disfavored, a consistent treatment of this scenario requires an appropriate rescaling of the mass priors, or of the prior of the models themselves, which are left for future work.

The mass ratio and the tidal deformabilities are sampled uniformly in $q \in [0.16,1]$ and $\Lambda_1, \Lambda_2 \in [0,5000]$, respectively. 
For the component spins, we take an aligned spin prior with the spin magnitudes in the range $[0, 0.90]$, while the luminosity distance is sampled uniformly in comoving volume in the range $[0, 500]$~Mpc\footnote{Spin magnitudes are usually assumed to be smaller for NS components, given that the Kepler limit implies $\chi \lesssim 0.7$. However, in order to analyze the data without assumptions about the source, one has to take into account that each component could also be a BH, thus with spin up to $\chi = 1$. The specific limit $\chi_{\rm max} = 0.90$ is due to waveforms validity constraints. This large range of spin magnitudes is then employed for all the models, even BNS approximants, in order to ensure a consistent prior volume.}.

\begin{figure}[!t]
        \centering
        \includegraphics[width=1.\linewidth]{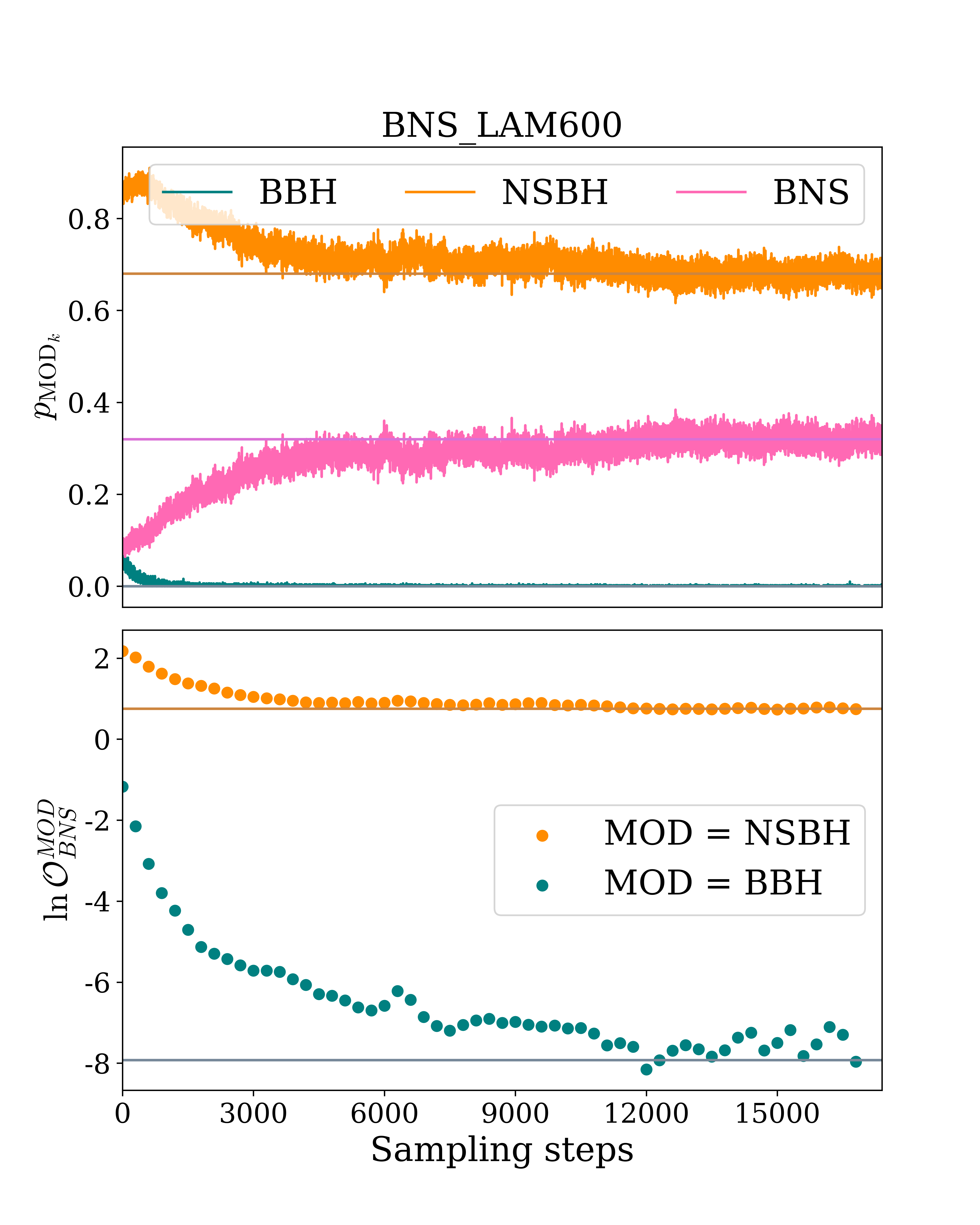}
    \caption{Top panel: Model probabilities, computed as the average number of samples over walkers in each model as a function of the sampling step for the \textsc{bns\_lam600} analysis (see Sec.~\ref{ssec:inj}). 
    Bottom panel: for the same analysis, odds ratios with respect to the injected model computed every 300 steps over the last 500 iterations.
    In both panels, the horizontal lines show the final values obtained in the last 500 iterations and discussed in Sec.~\ref{sec:results}.}
    \label{fig:convergence}
\end{figure}

\subsection{Convergence}
\label{sec:convergence}

The convergence of MCMC analyses is usually based on the accumulated number of effective samples, i.e., the samples thinned by the autocorrelation time, or on metrics such as the Gelman-Rubin coefficient~\cite{GRconv}. 
In the case of RJMCMC analysis, quantities such as  the autocorrelation time and the convergence statistics would need to be defined in a way that takes into account that each chain (walker) jumps between models and dimensions.

Trace plots are usually the best indicator of convergence for the parameter posteriors, and in the case of RJMCMC one can look whether the time spent in each model has become stable. Figure~\ref{fig:convergence} shows how the probabilities for each model and the odds ratios evolve over the sampler iterations, for the example of the \textsc{bns\_lam600} injection (see Sec.~\ref{ssec:inj} for details).

However, it is non-trivial to establish a definitive convergence criterion. 
Therefore, in this initial implementation, \texttt{t-roo} by default runs for a fixed number of steps and, if needed, the analysis can be resumed and run for more iterations, thanks to the implemented checkpoints.
Nonetheless, we also provide a few empirical criteria with which the sampler could automatically estimate convergence (see Appendix~\ref{sec:convergence_crit}).
In the future, we may investigate also alternative stopping criteria, more similar to standard MCMC analyses.

\section{Results: validation on simulated data and analysis of real events}
\label{sec:results}

\begin{table*}[htb]
\begingroup
\caption{Posterior model probabilities and odds ratios recovered by \texttt{t-roo} compared to the corresponding odds ratios obtained with \texttt{dynesty}, including their conservative uncertainty estimates.}
\label{tab:results_comp}
\renewcommand*{\arraystretch}{2}
\renewcommand\cellset{\renewcommand\arraystretch{1.3}}
\setlength{\tabcolsep}{8pt}
\begin{tabular}{l l l l}
        \hline
        & \multicolumn{1}{c}{$p_{\textsc{mod}}$ } &
        \multicolumn{2}{c}{ odds ratios } \\
        \hline 
        & \multicolumn{1}{c}{\texttt{t-roo}} & \multicolumn{1}{c}{\texttt{t-roo}}  & \multicolumn{1}{c}{\texttt{dynesty}}\\
        \hline
        \hline
        {\textsc{nsbh\_lam300}} & \makecell[tl]{$p_{\textsc{nsbh}} = 0.12$ \\ $p_{\textsc{bbh}}=0.88$} & $\ln \mathcal{O}_{\textsc{nsbh}}^\textsc{bbh} = 2.0$ & $ \ln \mathcal{O}_{\textsc{nsbh}}^\textsc{bbh} = 2.1 \pm 0.35 $ \\
                                            
        {\textsc{nsbh\_lam600}} & \makecell[tl]{$p_{\textsc{nsbh}} = 0.92$ \\ $p_{\textsc{bbh}}=0.083$}  & $\ln \mathcal{O}_{\textsc{nsbh}}^\textsc{bbh} = -2.4 $ & $\ln \mathcal{O}_{\textsc{nsbh}}^\textsc{bbh} =-1.9 \pm 0.35 $ \\

        {\textsc{nsbh\_lam1000}} & \makecell[tl]{$p_{\textsc{nsbh}} = 0.97$ \\ $p_{\textsc{bbh}}=0.033$}  & $\ln \mathcal{O}_{\textsc{nsbh}}^\textsc{bbh} = -3.4$ & $\ln \mathcal{O}_{\textsc{nsbh}}^\textsc{bbh} = -3.6 \pm 0.35 $ \\
        \textsc{bbh\_nsbhmasses} & \makecell[tl]{$p_{\textsc{nsbh}} = 0.24$ \\ $p_{\textsc{bbh}}=0.76$}  & $\ln \mathcal{O}_{\textsc{bbh}}^\textsc{nsbh} = -1.1 $ & $\ln \mathcal{O}_{\textsc{bbh}}^\textsc{nsbh} = -1.8 \pm 0.35$\\
        \textsc{bns\_lam600} & \makecell[tl]{$p_{\textsc{bns}} = 0.32$ \\ $p_{\textsc{nsbh}} = 0.68$ \\ $p_{\textsc{bbh}}=1.2\times 10^{-4}$}  & \makecell[tl]{$\ln \mathcal{O}_{\textsc{bns}}^\textsc{nsbh} = 0.75$ \\ $\ln \mathcal{O}_{\textsc{bns}}^\textsc{bbh} = -7.9$} & \makecell[tl]{$\ln \mathcal{O}_{\textsc{bns}}^\textsc{nsbh} = 2.0 \pm 0.34 $ \\ $\ln \mathcal{O}_{\textsc{bns}}^\textsc{bbh} = -6.8 \pm 0.34$} \\
        \textsc{bbh\_bnsmasses} & \makecell[tl]{$p_{\textsc{bns}} = 0.019$ \\ $p_{\textsc{nsbh}} = 0.083$ \\ $p_{\textsc{bbh}}=0.90$}  & \makecell[tl]{$\ln \mathcal{O}_{\textsc{bbh}}^\textsc{bns} = -3.8$ \\ $\ln \mathcal{O}_{\textsc{bbh}}^\textsc{nsbh} = -2.4$ } & \makecell[tl]{$\ln \mathcal{O}_{\textsc{bbh}}^\textsc{bns} = -9.3 \pm 0.34 $ \\ $\ln \mathcal{O}_{\textsc{bbh}}^\textsc{nsbh} = -4.0 \pm 0.33$ } \\
        \textsc{bns\_lam600\_mm} & \makecell[tl]{$p_{\textsc{phend}} = 1.2\times10^{-5}$ \\ $p_{\textsc{xas}}=4.8\times10^{-5}$ \\ $p_{\textsc{seob}}=3.2\times10^{-5}$ \\ $p_{\textsc{phendv2}} = 0.32$ \\ $p_{\textsc{xasv3}}=0.30$ \\ $p_{\textsc{seobv3}}=0.37$} & 
        \makecell[tl]{$\ln \mathcal{O}_{\textsc{xasv3}}^\textsc{phend} = -10 $ \\ $\ln \mathcal{O}_{\textsc{xasv3}}^\textsc{xas} = -8.8 $ \\ $\ln \mathcal{O}_{\textsc{xasv3}}^\textsc{seob} = -9.2$ \\ $\ln \mathcal{O}_{\textsc{xasv3}}^\textsc{phendv2} = 0.065$ \\ $\ln \mathcal{O}_{\textsc{xasv3}}^\textsc{seobv3} = 0.20 $ }&
        \makecell[tl]{$\ln \mathcal{O}_{\textsc{xasv3}}^\textsc{phend} = -11 \pm 0.32 $ \\ $\ln \mathcal{O}_{\textsc{xasv3}}^\textsc{xas} = -11 \pm 0.32 $ \\ $\ln \mathcal{O}_{\textsc{xasv3}}^\textsc{seob} = -11 \pm 0.32 $ \\ $\ln \mathcal{O}_{\textsc{xasv3}}^\textsc{phendv2} = -0.22 \pm 0.33 $ \\ $\ln \mathcal{O}_{\textsc{xasv3}}^\textsc{seobv3} = 0.43 \pm 0.32 $ } \\
        \textsc{nsbh\_lam600\_lowsnr} & \makecell[tl]{$p_{\textsc{nsbh}} = 0.70$ \\ $p_{\textsc{bbh}}=0.30$}  & $\ln \mathcal{O}_{\textsc{nsbh}}^\textsc{bbh} = -0.86 $ & $\ln \mathcal{O}_{\textsc{nsbh}}^\textsc{bbh} = 0.40 \pm 0.27 $ \\
        \textsc{nsbh\_lam300\_et} & \makecell[tl]{$p_{\textsc{nsbh}} = 0.99$ \\ $p_{\textsc{bbh}}=0.011$} & $\ln \mathcal{O}_{\textsc{nsbh}}^\textsc{bbh} = -4.5 $ & $\ln \mathcal{O}_{\textsc{nsbh}}^\textsc{bbh}-7.9 \pm 0.44 $\\
        \textsc{bns\_lam600\_et} & \makecell[tl]{$p_{\textsc{bns}} = 0.64$ \\ $p_{\textsc{nsbh}} = 0.36$ \\ $p_{\textsc{bbh}}=0.0$}  & \makecell[tl]{$\ln \mathcal{O}_{\textsc{bns}}^\textsc{nsbh} = -0.57$ \\ $\ln \mathcal{O}_{\textsc{bns}}^\textsc{bbh} = -\infty$ }& \makecell[tl]{$\ln \mathcal{O}_{\textsc{bns}}^\textsc{nsbh} = -1.9 \pm 0.38 $ \\ $\ln \mathcal{O}_{\textsc{bns}}^\textsc{bbh} = -217 \pm 0.37 $}     \\
        
        \hline

\end{tabular}
\endgroup
\end{table*}

To assess the performance of \texttt{t-roo}, we compare the results from the injection runs to an equivalent set of analyses performed with \textsc{bilby} and the \texttt{dynesty} sampler~\cite{dynesty}, both for 
parameter posteriors and model selection.

Usually, the odds ratio between two hypotheses, or models, $\mathcal{H}_B$ and $\mathcal{H}_A$ is computed as
\begin{equation}
    \mathcal{O}^B_A = \frac{\mathcal{Z}_{\mathcal{H}_B}}{\mathcal{Z}_{\mathcal{H}_A}} \equiv \frac{p(\mathcal{H}_B | d)}{p(\mathcal{H}_A | d)} = \frac{p(d | \mathcal{H}_B)}{p(d| \mathcal{H}_A)} \frac{p(\mathcal{H}_B)}{p(\mathcal{H}_A)}.
    \label{eq:odds_ev}
\end{equation}
If we assume no prior knowledge about the model preferences, this reduces to the \emph{Bayes factor} $\mathcal{B}_A^B$, i.e., to the ratio of evidences $\mathcal{Z}_{\mathcal{H}_B}/\mathcal{Z}_{\mathcal{H}_A}$. 
As mentioned earlier, for RJMCM samplers, the odds ratio is given by the percentage of samples in a given model, i.e.:
\begin{equation}
    \mathcal{O}_A^B = \frac{n_B}{n_{\rm tot}} \frac{n_{\rm tot}}{n_A} = \frac{n_B}{n_A}.
    \label{eq:odds_rj}
\end{equation}

Table~\ref{tab:results_comp} lists the model probabilities and odds ratios, computed using Eq.~\eqref{eq:odds_rj}, found with \texttt{t-roo} and the corresponding odds ratios from the \texttt{dynesty} runs, evaluated with Eq.~\eqref{eq:odds_ev}, for comparison. 
Model probabilities in \texttt{t-roo} are computed by taking the total number of samples over walkers in each model during the last 500 sampling steps, and odds ratios are always shown with respect to the injected model.

Overall, in most cases we find good consistency between odds ratios obtained with \texttt{t-roo} and \texttt{dynesty}. 
\texttt{dynesty} provides an uncertainty estimate for the computed evidence $\sigma(\ln \mathcal{Z})$, quantified from both statistical and sampling error contributions~\cite{dynesty}.
In Table~\ref{tab:results_comp}, next to the \texttt{dynesty} estimates of the odds ratio between two models $\textsc{mod}_1$ and $\textsc{mod}_2$, we report the uncertainty\footnote{This is a conservative estimate that effectively ignores potential correlations between evidences on the two models.}
\begin{equation}
    \sigma \left( \ln \mathcal{O}_{\textsc{mod}_1} ^{\textsc{mod}_2}\right) \sim \sigma \left( \ln \mathcal{Z}_{\textsc{mod}_1}\right) + \sigma \left( \ln \mathcal{Z}_{\textsc{mod}_2}\right).
\end{equation}

\begin{figure*}[tb]
        \centering
        \includegraphics[width=0.7\linewidth]{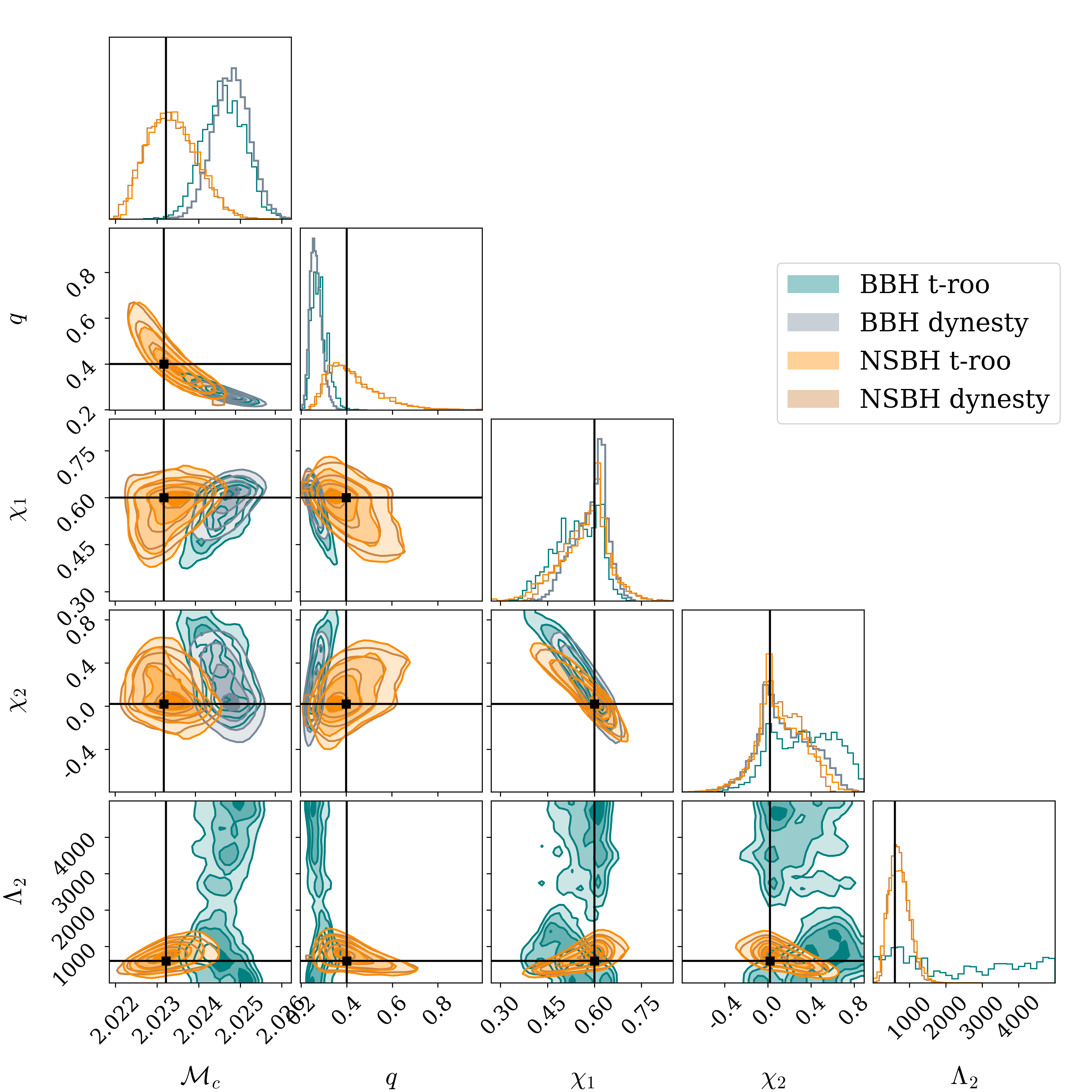}
    \caption{Parameter posterior distributions recovered by \texttt{t-roo} for both the NSBH and BBH models for the \textsc{nsbh\_lam600} injection, compared to the posteriors obtained for the same analysis with \texttt{dynesty} and \textsc{Bilby}. For \texttt{t-roo}, we show only the samples of the last 1000 sampling steps to exclude samples where the sampler is still exploring the preferred model. The black lines indicate the injected values.}
    \label{fig:corner_nsbh600}
\end{figure*}

For the odds ratios recovered with RJMCMC samplers, we cannot estimate uncertainty in the same way, since the models' probabilities derive as a consequence of the sampler visiting more often the more favored models\footnote{To provide some uncertainty, one could compute the probability of each model as the mean number of samples in that model over the walkers. This value would be characterized by the statistical uncertainty $\sigma = 1/\sqrt{N_w}$, with $N_w$ being the number of walkers considered. However, given the large number of walkers employed (specifically, $N_w = 500$ in all the analyses presented here), this uncertainty would still be negligible compared to the \texttt{dynesty} one.}. 

However, once the sampler is converged, we expect a good agreement with the odds ratio recovered by \texttt{dynesty}. Nevertheless, we do not expect to recover the exact same value obtained with \texttt{dynesty}, given the differences between the samplers and that also the evidences recovered with nested samplers can fluctuate; see Appendix~\ref{sec:evidences} for a more extended discussion.

As an important validation, in all the tests, the model preferred by \texttt{t-roo} and \texttt{dynesty} is the same.
We find larger differences in the recovered odds ratios in the cases where we compare more than two models. This is likely due to the fact that the evidence, as computed in \texttt{dynesty} and shown in Eq.~\ref{eq:evidence}, involves a marginalization over the entire parameter space. In \texttt{t-roo}, the less favored models are visited less often, and consequently also the parameter space is less explored. 
To verify that, instead, this is not caused by some issue within the sampler, we re-analyze the injection \textsc{bns\_lam600} with \texttt{t-roo} using for recovery only the NSBH and BNS models. If the sampler works properly, we expect to recover the same model probabilities, and consequently odds ratio $\ln \mathcal{O}_\textsc{bns}^{\textsc{nsbh}}$, as in the analysis including also the BBH model, because, once converged, a RJMCMC sampler should return the ``correct" model probabilities, independently of the number of models investigated. With this new analysis, we obtain $p_\textsc{nsbh}=0.68$ and $p_\textsc{bns}=0.32$, yielding $\ln \mathcal{O}_\textsc{bns}^{\textsc{nsbh}}=0.78$, which correspond to the values reported in Table~\ref{tab:results_comp} for the analysis including also the BBH model (small differences in the odds ratio value are due to fluctuations in the next significant digits in the model probabilities). Therefore, we confirm that the differences with the \texttt{dynesty} results are not due to issues in the sampler, but to the fact that for the least favored models the collected samples do not cover the full parameter space.
In order to recover the same evidences, and consequently odds ratio, one could run the analysis for longer and potentially tweak the priors to ensure a full exploration of all the parameter spaces. However, this would undermine the computational advantage of \texttt{t-roo} and would be in contrast to the wish of sampling more preferred models more often, while avoiding unnecessary evaluations of the less favored ones.

\subsection{Injection-recoveries with NSBH and BBH models}

Figure~\ref{fig:corner_nsbh600} shows the recovered posteriors for the intrinsic parameters $\mathcal{M}_c$, $q$, $\chi_1$, $\chi_2$ and $\Lambda_2$ for the \textsc{nsbh\_lam600} analysis, comparing the posteriors obtained both for \texttt{t-roo} and \texttt{dynesty} with the NSBH model \texttt{IMRPhenomNSBH} and the BBH one \texttt{IMRPhenomD}. 
The injected parameters are recovered correctly for the NSBH approximant and slightly biased for the recovery with the BBH one, as expected given that the former was used to inject the signal. 
In particular, the separated peaks of the $\mathcal{M}_c$ and $q$ posteriors illustrate why we need specific proposals in the between-model move, as explained in Sec.~\ref{sec:mass_proposal}. 
The posteriors recovered by \texttt{t-roo} are consistent with the ones obtained from \textsc{Bilby} using \texttt{dynesty} for both models.
The \texttt{t-roo} posterior for the BBH model appears somewhat rugged, as a consequence of the BBH model being disfavored and therefore rarely visited by the walkers.
In fact, this shows the advantage of RJMCMC samplers and how they can reduce the computational cost of these analyses when one of the models is strongly preferred, as the sampler will then focus automatically on the model with the higher preference.

The marginalized distribution for $\Lambda_2$ shows how \texttt{t-roo} accurately recovers the right value in the NSBH model, whereas in the BBH model $\Lambda_2$ is distributed across the entire prior range thanks to the random walk proposal. 
In the absence of tidal information, the BBH model would be preferred and the sampler would spend more time there; consequently, the large number of iterations of the random walk would consistently yield a final posterior that just recovers the prior on $\Lambda_2$.

For the NSBH injections \textsc{nsbh\_lam600} and \textsc{nsbh\_lam1000},
\texttt{t-roo} correctly recovers \texttt{IMRPhenomNSBH} as the favored model,
while for \textsc{nsbh\_lam300}, the sampler favors the BBH approximant \texttt{IMRPhenomD}.
This is expected, because the tidal information recovered by the data in the latter case is not sufficient to discriminate between the two scenarios, and the sampler prefers the simpler model. 
Notably, this is consistent with the \texttt{dynesty} evidences, which also indicate a preference for the BBH model for this particular injection. 

On the other hand, for \textsc{nsbh\_lam600\_lowsnr}, i.e., the injection with $\Lambda_2 = 600$ but placed at a larger distance, we still find that the NSBH model is favored, but only with $p_{\textsc{nsbh}} = 70\%$ compared to $p_{\textsc{nsbh}}=92\%$ for the injection at $d_L=30$~Mpc. 
Also these results are consistent with the ones from \texttt{dynesty}.

The situation changes when we analyze the signal from the same source but detected by ET: thanks to the larger SNR ($\sim 519$), also a smaller tidal deformability is enough to clearly recover the tidal content, and therefore discriminate between the two scenarios. 
For the \textsc{nsbh\_lam300\_et} test, both \texttt{t-roo} and \texttt{dynesty} correctly identify \texttt{IMRPhenomNSBH} as the preferred model.

\begin{figure}[t]
        \centering
        \includegraphics[width=1.\linewidth]{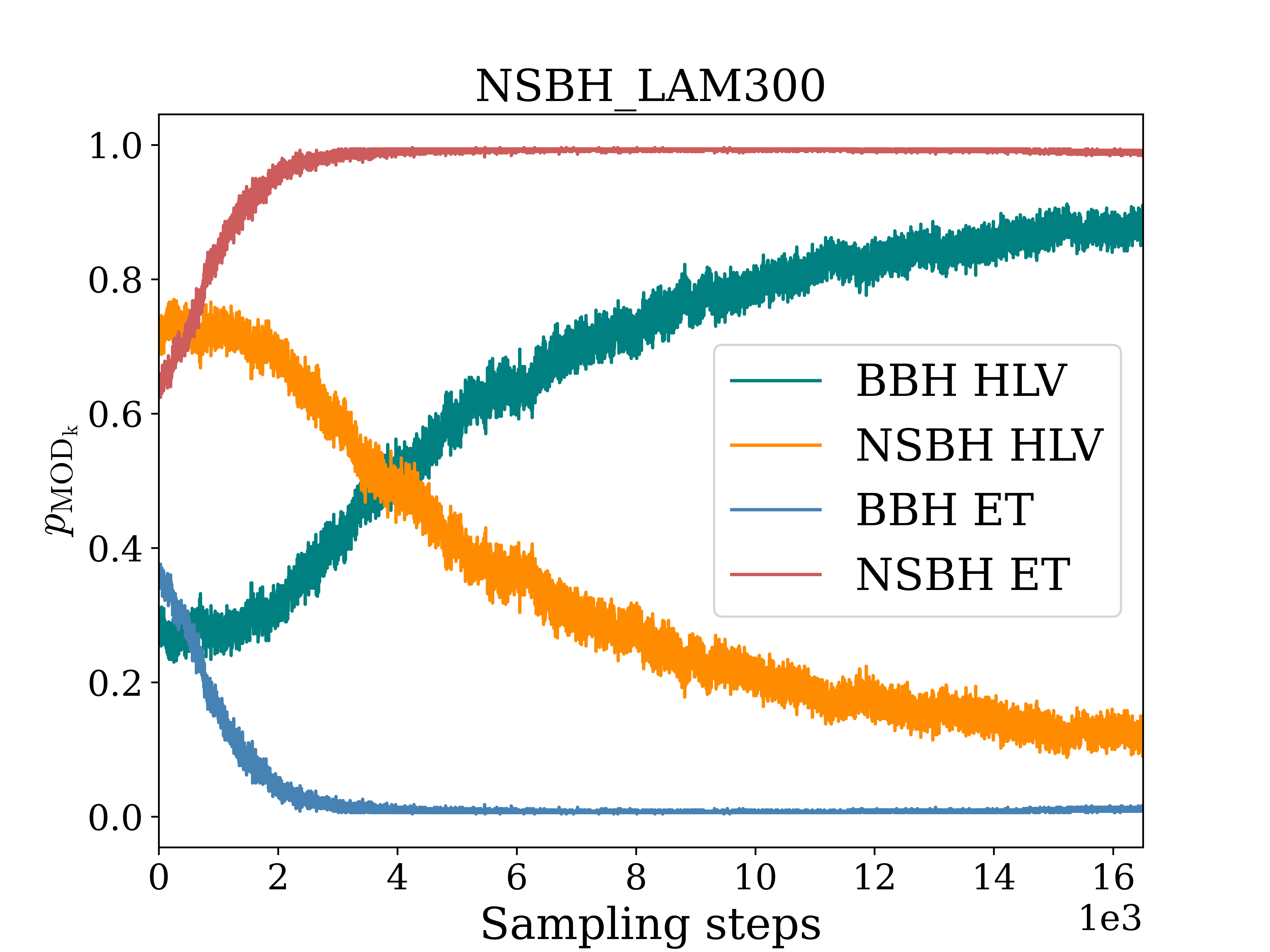}
    \caption{Model probabilities for the \textsc{nsbh\_lam300} injection as recovered by a LIGO-VIRGO and an ET network (labeled ``HLV" and ``ET", respectively), computed as the average number of samples per walker accumulated in each model over the RJMCMC sampling steps. The model probabilities do not start exactly at 0.5 because of the temperature swaps in the preliminary sampling.}
    \label{fig:conv_comp}
\end{figure}

Moreover, we note how, for such high-SNR signals, the data are very informative and therefore \texttt{t-roo} manages to converge very quickly, as shown in Fig.~\ref{fig:conv_comp}. 
This highlights the potential of RJMCMC algorithms in these scenarios: large SNRs usually make parameter estimation analyses more computationally expensive, and in these cases running separate analyses with all the models could become prohibitive. On the other hand, with \texttt{t-roo} large SNRs make it easier for the sampler to understand what the favored model is, yielding a significant computational advantage.

If, instead, we simulate a BBH signal with the same parameters (i.e., the analysis \textsc{bbh\_nsbhmasses} in Table~\ref{tab:injections}), we see from Table~\ref{tab:results_comp} that \texttt{t-roo} correctly recovers the BBH model as the preferred one, with an odds ratio again consistent with the \texttt{dynesty} one. 
Also in this case, the inferred posteriors are in excellent agreement with the ones from \textsc{Bilby} and the $\Lambda_2$ parameter for the NSBH model rails against zero, indicating that no tidal information is recovered in the data.

\begin{figure*}[tb]
        \centering
        \includegraphics[width=0.8\linewidth]{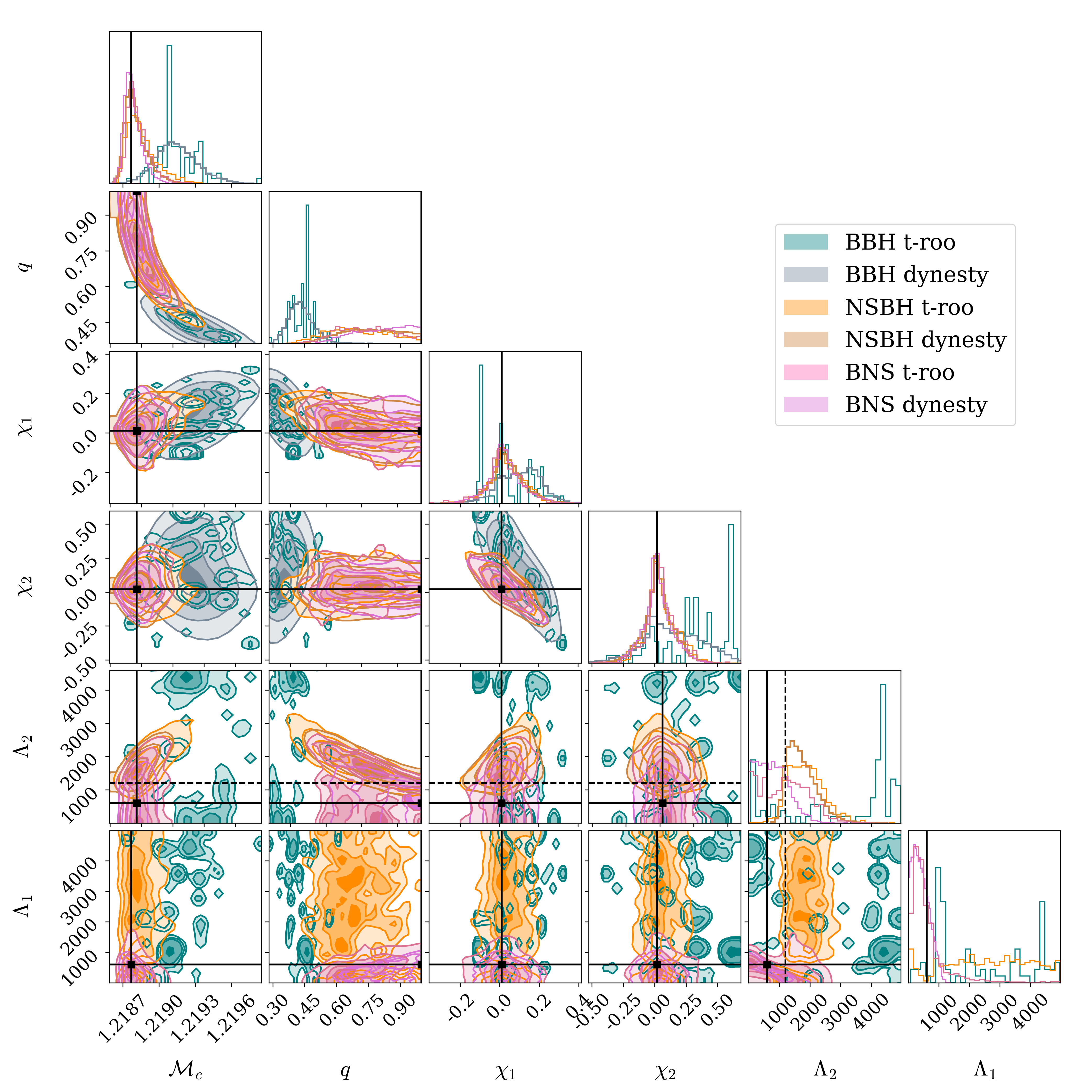}
    \caption{Parameter posterior distributions recovered by \texttt{t-roo} with BNS, NSBH, and BBH models for the \textsc{bns\_lam600} injection, compared to the posteriors obtained for the same analysis with \texttt{dynesty}. 
    For \texttt{t-roo}, we show only the samples of the last 1000 sampling steps, to avoid including samples of a phase when the sampler is still exploring the preferred model. The black solid lines indicate the injected values, while the dashed line in the panel for the $\Lambda_2$ one-dimensional distribution denotes the $\Lambda_2$ value needed by the NSBH model to yield the same $\tilde{\Lambda}$ value as the injected one.}
    \label{fig:corner_bns600}
\end{figure*}

\subsection{Injection-recoveries with BBH, NSBH, and BNS models}

In Fig.~\ref{fig:corner_bns600}, we show the posteriors recovered for the BNS injection \textsc{bns\_lam600}. 
We used a BBH, NSBH, and BNS model for the recovery, as listed in Table~\ref{tab:injections}. 
We find again very good agreement between the distributions inferred with \texttt{t-roo} and \texttt{dynesty}. 
The injected values are correctly recovered by the BNS and NSBH models, while we find biases in the BBH ones, especially in chirp mass and mass ratio, where the latter peaks at $q=0.44$ although the injected signal corresponds to an equal-mass system.
This is expected given the lack of tidal content in the BBH model and the correlation between the tidal parameters and mass ratio. 

The \text{t-roo} posterior for the BBH model is rather scattered, because it encompasses less than $0.1\%$ of the total number of samples, due to its low preference.
We highlight that this is the expected outcome of RJMCMC samplers: the walkers will systematically explore mostly the favored model(s), therefore yielding a good estimate of their posteriors. Posteriors for the disfavored models could be obtained by running the sampler for longer, until the accumulated number of samples in the disfavored model is enough to reconstruct the parameters posterior probability. However, this undermines the purpose of RJMCMC, i.e., understanding which models are more likely and focusing on them. The underlying idea is that a RJMCMC sampler will provide us with an estimate of the preferred model and the ``preferred'' posterior probabilities, i.e., the posteriors conditional to the model deemed ``correct''.

From the models' probabilities in Table~\ref{tab:results_comp}, the favored model is \texttt{IMRPhenomNSBH}, even though the injected signal was simulated with the BNS approximant \texttt{IMRPhenomD\_NRTidalv2}. 
This is again in agreement with the odds ratios estimated by \texttt{dynesty}. 
Looking at the marginalized posterior for $\Lambda_2$ in Fig.~\ref{fig:corner_bns600}, we see that for the NSBH model both \texttt{t-roo} and \texttt{dynetsy} recover a value larger than the injected one, $\Lambda_{2, \rm inj} = 600$, peaking roughly at $\Lambda_2 \sim 1800$. 
At leading order, the tidal information enters waveform approximants as the mass-weighted combination $\tilde{\Lambda}$~\cite{Flanagan:2007ix, Wade:2014vqa}, which, in the case of the \textsc{bns\_lam600} injection, is $\tilde{\Lambda} = 600$. 
When using the NSBH model for this specific system, we would need a NS tidal deformability of  $\Lambda_2 = 1200$ to yield the injected value of $\tilde{\Lambda}$, since the tidal deformability of the BH is set to zero.
Therefore, the NSBH model is trying to recover the tidal information, but assuming it all belongs to just one object. 
This also explains why the NSBH model is preferred: since the tidal content at this point is basically equivalent in the BNS and NSBH models, in the absence of other discriminant information the sampler prefers the simpler model. 
This means that in principle we cannot distinguish signals from a BNS or NSBH source just based on the leading-order tidal information in different approximants, but this is not caused by issues in the specific sampling methods used.
Employing more advanced models, such as \texttt{NRTidalv3}~\cite{Abac:2023ujg}, which can distinguish the individual component's tidal deformabilities, will help to solve this degeneracy.

Because of the large number of samples accumulated for the NSBH model, its posterior for $\Lambda_1$ shows how the sampler recovers the uniform prior for this \textit{pseudo}-parameter that, as explained in Sec.~\ref{sec:proposals}, basically serves as a storage variable.

As for Fig.~\ref{fig:corner_nsbh600}, we see a noticeable difference in the posterior peaks of $\mathcal{M}_c$ and $q$ between the BBH model and the NSBH or BNS one. 
Considering that, as discussed above, the NSBH and BNS models recover the same tidal information, this difference arises likely from the lack of tidal content in the BBH approximant, and supports the need of a tidal-dependent proposal for these two parameters.

Finally, for the \textsc{bbh\_bnsmasses} injection,  where we analyze a signal coming from a source with the same parameters, but being a BBH system instead of a BNS one, \texttt{t-roo} correctly recovers \texttt{IMRPhenomD} as the preferred model.
In this case, the mass ratio is recovered correctly by all models, since the injection is performed with a BBH approximant (i.e., the data do not contain tidal information), and both the NSBH and BNS models recover correctly $\Lambda_1$ and/or $\Lambda_2$ being zero.

\begin{figure*}[tb]
        \centering
        \includegraphics[width=0.8\linewidth]{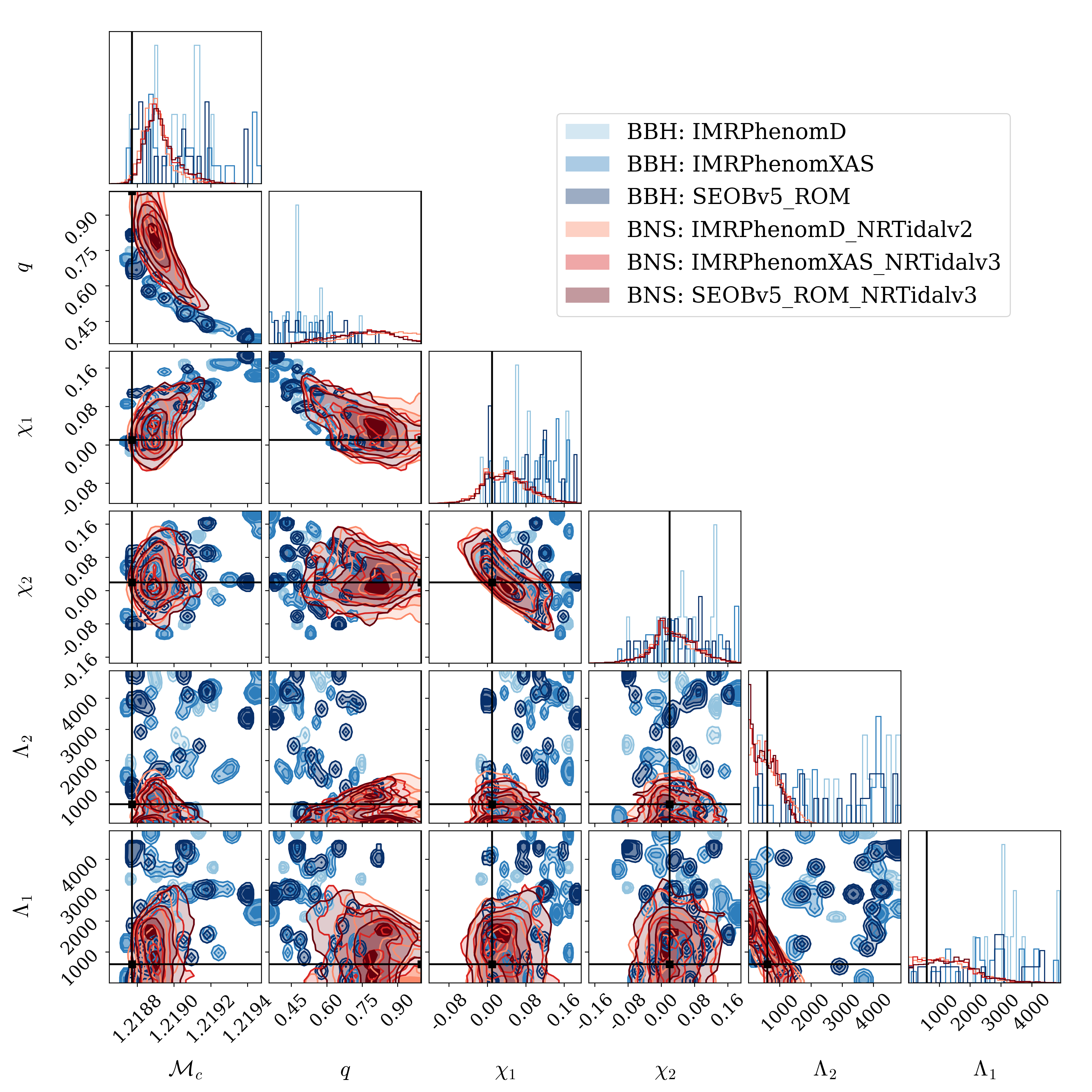}
    \caption{Parameter posterior distributions recovered by \texttt{t-roo} with three different BNS and three different BBH models for the \textsc{bns\_lam600\_mm} injection. To better show the comparison of the posteriors for the BBH models, in this case we include in the corner plot the last 1500 iterations. The black solid lines correspond to the injected values.}
    \label{fig:corner_moremodels}
\end{figure*}

\subsection{Injection comparing multiple models}

Finally, we discuss the case in which we compare not only one model per binary type in the recovery, but multiple models describing the same source. 
For \textsc{bns\_lam600\_mm}, we use \texttt{IMRPhenomXAS\_NRTidalv3} to inject a signal from an equal-mass BNS system with $\Lambda_1 = \Lambda_2 = 600$, and analyze it with three BNS and three BBH approximants; cf. Table~\ref{tab:injections}. 

Figure~\ref{fig:corner_moremodels} shows the posteriors recovered with the different models. 
Although not shown here for simplicity, we find in all cases a very good agreement with the posteriors recovered by \textsc{bilby} runs with \texttt{dynesty}. 
We only see slight differences in the $q$ posteriors recovered by BNS models. 
This is likely due to the interplay with the BBH model, which systematically prefers more unequal-mass systems and slows down the convergence for this parameter: a more sophisticated proposal for the mass ratio, potentially also including spin information, would help mitigate this effect. 

We see that all BBH approximants recover very similar posteriors, and likewise all posteriors from BNS models are essentially identical.
However, there are noticeable differences between BBH and BNS posteriors, especially for $\mathcal{M}_c$ and $q$, in analogy to the previous results. 
The BNS models recover $\Lambda_1$ and $\Lambda_2$ correctly, while the BBH ones tend to return the prior. 
\texttt{t-roo} correctly favors the BNS models, and in particular \texttt{SEOBNRv5\_ROM\_NRTidalv3}. 
Moreover, although the odds ratios are slightly different, the order of the model preference, even within the same kind of source, is the same as \texttt{dynesty}. The only difference is that \texttt{t-roo} slightly prefers \texttt{IMRPhenomD\_NRTidalv2} over \texttt{IMRPhenomXAS\_NRTidalv3}. However, the difference with respect to the \texttt{dynesty} odds ratio lies within the uncertainty of the latter. 

We remark that the \texttt{t-roo} results are obtained with only one run, in which we can compare an arbitrary number of models for different kinds of sources. 
Even though a larger set of models might require a larger number of iterations for the sampler to reach convergence, this is still computationally and practically more convenient than running separate analyses for each model, especially if some of the models, like the BBH ones here, are quite strongly disfavored.

We emphasize again that \texttt{t-roo} and RJMCMC samplers focus on model selection, returning a good posterior estimation only if a model is sufficiently favored. Basically, the sampler does not ``waste'' time exploring the posteriors for models that it has inferred do not describe the data sufficiently well, which yields a significant computational gain.

\begin{table}[tb]
\begingroup
\renewcommand*{\arraystretch}{2}
\setlength{\tabcolsep}{6pt}
\begin{tabular}{l c c c}
        \hline
        Event & $p_{\textsc{bbh}}$ & $p_\textsc{nsbh}$ & $p_\textsc{bns}$ \\
        \hline

        GW190425 &  0.66 & 0.31 & 0.033 \\
        GW230529 &  0.38 & 0.55 & 0.067 \\
   
       \hline

\end{tabular}
\caption{\texttt{t-roo} recovered probabilities for models corresponding to different source types for the real events GW190425 and GW230529.}
\label{tab:real_events}
\endgroup
\end{table}

\subsection{Real events}

We employ \texttt{t-roo} to analyze the real events GW190425 and GW230529\footnote{Data were downloaded from the Gravitational Wave Open Science Center~\cite{gw190425_data, gw230529_data}, with the estimated power spectral densities provided in Refs.~\cite{gw190425_psd, gw230529_psd}, respectively}, 
for which the absence of electromagnetic counterparts and of clear tidal signatures did not enable a certain classification of the components as NSs.
We do not analyze GW190814 because, given the large mass ratio,
higher-order-modes would need to be included in all models used to analyze this signal\footnote{Recently, an NSBH model including higher-order-modes has been developed in Ref.~\cite{RamisVidal:2026ycb}. However, for this work we employ only publicly available models reviewed by the LVK collaboration.}.

We analyze both events with \texttt{IMRPhenomXAS} for the BBH case, \texttt{IMRPhenomNSBH} for the NSBH case, and \texttt{IMRPhenomXAS\_NRTidalv3} for the BNS one. 
The recovered probabilities for each source are shown in Table~\ref{tab:real_events}. 
For GW190425, \texttt{t-roo} prefers the BBH model, with a probability $p_{\textsc{bbh}} = 0.66$, consistent with the fact that no significant tidal information was recovered for this event. 
Among the NSBH and BNS models, the NSBH one is preferred, again consistently with the fact that it has fewer free parameters. 

For GW230529, \texttt{t-roo} finds that the NSBH model \texttt{IMRPhenomNSBH} is the one that best describes the data, with a probability $p_{\textsc{nsbh}} = 0.55$.

\subsection{Computational cost}
\label{ssec:computational}

One of the main advantages of RJMCMC algorithms for Bayesian model selection is that, especially when the data are informative, the computational cost is reduced with respect to performing separate analysis for each model and then comparing the evidences. 

A consistent comparison between the samplers' efficiency is beyond the scope of this paper because it would require specific prescriptions to describe computational costs that take into account not only the number of accumulated samples and the likelihood evaluation time, but also the parallelization inefficiency, the burn-in time, and the effect of parallel tempering, as shown, for example, in Ref.~\cite{Ashton:2021anp} for \textsc{Bilby}-MCMC\footnote{As \texttt{dynesty} is a nested sampler, a different model for the computational cost estimation would be needed}.

To provide an indication of the computational gain, the \texttt{t-roo} analysis for \textsc{nsbh\_lam600} performed the preliminary sampling stage and accumulated $10^4$ RJMCMC samples in 14 hours, while for \texttt{dynesty} the BBH run took a bit more than 24 hours, accumulating 10159 samples,
and the NSBH one took $\sim 31$ hours, accumulating 10889 samples. 
All these times refer to runs performed with 8 cores on the same machine.
Even though \texttt{eryn}, and consequently \texttt{t-roo}, is parallelized, the current implementation is suboptimal, as only the likelihood evaluations are performed on different physical CPUs, while every other operation is executed on one main device, leading to a significant overhead cost. 
Consequently, \texttt{eryn} does not scale properly when running on more than 8 cores. 
\texttt{dynesty}, on the other hand, can run on many CPUs in parallel, making it faster in practice. 
In order to fully exploit the potential of \texttt{t-roo}, and especially to make it competitive with samplers like \texttt{dynesty} that can run in parallel on many more cores, the code parallelization has to be optimized.

\section{Conclusions and future applications}
\label{sec:conclusions}

We have developed \texttt{t-roo}, the first RJMCMC sampler to analyze GW signals from CBCs with different state-of-the-art waveform models, even with a different number of parameters, at the same time.
Our implementation is focused towards applying this method to source classification, i.e., comparing models describing different types of binary systems such as BNSs, BBHs, and NSBHs. 
We implement specific between- and within-model proposals that handle the different tidal content of the models. 
Moreover, we encompass the possibility of comparing multiple different models for each of various source types at the same time. 

We assessed the performance of \texttt{t-roo} in different scenarios, analyzing simulated GW detections from different sources (see Sec.~\ref{sec:results} for details). 
In all the cases, the inferred posterior distributions and odds ratios are in agreement with the ones obtained from \textsc{Bilby} analyses with the \texttt{dynesty} sampler. 

For NSBH signals, \texttt{t-roo} correctly identifies the binary type when the tidal information in the data is sufficient to discriminate the two options, otherwise it prefers the simpler BBH model. 
This means that signals from systems with a lower tidal deformability require larger SNR to be correctly identified, as expected.
However, an NSBH system with $\Lambda_2 = 600$ is correctly identified already at an SNR of 14.
On the other hand, when injecting BNS signals, the NSBH model is favored, due to the fact that it can mimic the tidal content of the injection, while requiring one fewer free parameter.
However, in all cases, the models preferred by \texttt{t-roo} are consistent with the \texttt{dynesty} results.
We also investigated BBH signals from sources where masses are within the BNS or NSBH range, finding that the BBH model is correctly preferred in both cases. 
Finally, we analyzed a BNS injection with three BNS and three BBH models, finding again that the inferred odds ratios and posteriors agree with the ones recovered by \texttt{dynesty}.

We emphasize that the main advantage of an RJMCMC approach is that it only requires one run to provide model probabilities and posteriors for the models sufficiently favored, while otherwise one run for each model is needed. 
It can significantly reduce the computational cost of the analysis, because, the less favored a model is, the less time the sampler will spend in that model; however, \texttt{t-roo} specifically needs to be better parallelized in order to fully exploit its computational-gain potential.

This reduction of the computational cost becomes more pronounced when the data is particular informative, and when we are comparing many models.
The advantage of performing only one run is not only computational, but also practical, since it reduces the human overhead, the setup time, and the potential risk of mistakes or mismatches in settings between runs.

Further developments require, first of all, an improvement of the code parallelization, together with investigating robust convergence criteria and potential methods to estimate uncertainties.
Moreover, future work will not just focus on the improvements mentioned throughout the work, but also on the application to different case studies. 
Given the flexibility of the framework, different use-cases simply require the design of specific proposals, in order to make the RJMCMC jumps efficient. 
For example, it will be interesting to compare models with precession and eccentricity, or models with or without deviations from GR described by extra parameters.

 The computational gain of this approach is going to be particularly important with the observation of long and loud signals. 
 While the analysis of high-SNR signals is usually more computationally expensive, for RJMCMC samplers it becomes easier to distinguish the information, and consequently it requires fewer iterations to converge.
 Therefore, \texttt{t-roo} is a very powerful starting tool for a large number of applications in GW data analysis not only for current detectors, but also for future ones such as the third-generation ground-based detectors Einstein Telescope and Cosmic Explorer, and the space-based LISA.

\section*{Acknowledgments}

We thank Alexandre Toubiana and Alessandro Santini for the help with \texttt{eryn}, and Nihar Gupte for the useful comments.
A.P., T.D., and H.K. acknowledge funding from the EU Horizon under ERC Starting Grant, no.\ SMArt-101076369. Views and opinions expressed are those of the authors only and do not necessarily reflect those of the European Union or the European Research Council. Neither the European Union nor the granting authority can be held responsible for them. JG acknowledges support from the Simons Foundation International via Grant No. SFI-MPS- BH-00012593-06. 
Computations were performed on the DFG-funded research cluster Jarvis at the University of Potsdam (INST336/173-1; project number: 502227537).
This material is based upon work supported by NSF's LIGO Laboratory which is a major facility fully funded by the National Science Foundation.

\appendix

\onecolumngrid

\section{Reversible Jump MCMC theory}
\label{sec:rjmcmc_theory}

\subsection{Bayesian model selection}

According to Bayes' theorem, we can infer the parameters $\vec{\theta}$ of an observed system from the observed data $d$ as
\begin{equation}
p(\vec{\theta}|d,\mathcal{H}) = \frac{p(d | \vec{\theta}, \mathcal{H}) p (\vec{\theta}|\mathcal{H})}{p(d|\mathcal{H})},
\end{equation}
where $p(\vec{\theta}|d,\mathcal{H})$ is the posterior probability of the parameters $\vec{\theta}$ given the data $d$ and a model or hypothesis $\mathcal{H}$ to describe the data, $p(d | \vec{\theta}, \mathcal{H})$ is the \textit{likelihood}, and $p(\vec{\theta}|\mathcal{H})$ is the prior probability of the parameters under the assumption of the model or hypothesis $\mathcal{H}$, i.e., the prior describes our knowledge about the parameters before the observation.

The term at the denominator is called \textit{evidence} and corresponds to the marginalized likelihood
\begin{equation}
    p(d|\mathcal{H}) = \int d\vec{\theta} p(d | \mathcal{H}, \vec{\theta}) p (\vec{\theta} | \mathcal{H}). 
    \label{eq:evidence}
\end{equation}
When interested in the parameters posterior probabilities $p(\vec{\theta}|d,\mathcal{H})$, the evidence serves just as a normalization factor. 
However, it plays a crucial role when the goal is model selection instead.

Let us suppose that we have two models $\mathcal{H}_\textsc{a}$ and $\mathcal{H}_\textsc{b}$, and we want to know which one provides a better description of some observed data $d$. Using Bayes theorem, the probability of each model given the data is
\begin{equation}
    p(\mathcal{H}_\textsc{a} | d) = \frac{p(\mathcal{H}_\textsc{a}) p(d | \mathcal{H}_\textsc{a}) }{p(d)},
\end{equation}
and similarly for $\mathcal{H}_\textsc{b}$. To compare the models' probabilities, one usually refers to the odds ratio
\begin{equation}
    \mathcal{O}_\textsc{a}^\textsc{b} = \frac{p(\mathcal{H}_\textsc{b} | d)}{p(\mathcal{H}_\textsc{a} | d)} = 
    \underbrace{\frac{p(\mathcal{H}_\textsc{b})}{p(\mathcal{H}_\textsc{a})}}_{\pi_\textsc{a}^\textsc{b}} \underbrace{\frac{p (d | \mathcal{H}_\textsc{b})}{ p (d | \mathcal{H}_\textsc{a})}}_{\mathcal{B}_\textsc{a}^\textsc{b}} \frac{\cancel{p(d)}}{\cancel{p(d)}},
\end{equation}
where the ratio of the models' prior probabilities $\pi_\textsc{a}^\textsc{b}$ is usually set to 1, meaning that we do not have a priori preference between them. Therefore, usually the odds ratio reduces to the \textit{Bayes factor} $\mathcal{B}_\textsc{a}^\textsc{b}$, defined as the ratio between the evidences obtained with the different models, computed as in Eq.~\eqref{eq:evidence}.

An alternative approach consists in sampling directly over the model, i.e., the target of the Bayesian inference becomes the joint posterior probability
\begin{equation}
    p(\vec{\theta}_k, k | d) = \frac{p(d | k, \vec{\theta}_k) p(k, \vec{\theta}_k) }{\sum_{k' \in K} \int p(k', \vec{\theta}'_{k'}) p(d | k', \vec{\theta}'_{k'}) d \vec{\theta}'_{k'}},
    \label{eq:bayes_mm}
\end{equation}
with $k$ being the model's label for a finite set $K$ of models, $\vec{\theta}_k$ the parameters of the specific model $k$, and $p(k,\vec{\theta}_k) = p(k)p(\vec{\theta}_k | k)$ the prior for each model and its parameters.
This joint posterior probabilities can always be factorized in
\begin{equation}
    p(k,\vec{\theta}_k | d) = p(k | d) p(\vec{\theta}_k | k, d),
\end{equation}
hence providing the model's probability given the data and the parameters probability for a given model, given the data and a specific choice of model.

Reversible Jump MCMC algorithms sample directly the joint posterior $p(\vec{\theta}_k, k | d)$. The key idea is that, the more (less) favored a model is, the more (less) time the sampler will spend in it, and consequently the more (less) samples it will accumulated for that model.

Therefore, at the end of a RJMCMC algorithm, the odds ratio between two models $\mathcal{H}_\textsc{a}$ and $\mathcal{H}_\textsc{b}$ will be given by
\begin{equation}
    \mathcal{O}_\textsc{a}^\textsc{b} = \frac{n_\textsc{b}}{n_\textsc{a}},
\end{equation}
with $n_\textsc{a}$ and $n_\textsc{b}$ being the number of samples in model $\mathcal{H}_\textsc{a}$ and $\mathcal{H}_\textsc{b}$, respectively.

\subsection{MCMC review}

Markov chains are stochastic processes that define a sequence of states of a system, in which each state depends on the previous one, and only on that. 
A Markov chain is characterized by a transition kernel $\mathcal{T}$, which describes the transition probability from one state to the other --- i.e., $\mathcal{T}(x,y)$ denotes the conditional probability of the next state in the chain to be $y$ given that the present state is $x$---, and a limiting distribution $\pi$, to which the states probabilities stabilize after an infinite time, or an infinite number of steps.
The limiting distribution of a Markov chain is stationary, i.e.,  it does not change over time, or $\pi \mathcal{T} = \pi$.

If a Markov chain is \textit{irreducible}, i.e., each state can be reached from any other state in a finite,  however large, number of steps, and \textit{aperiodic}, i.e., we can return to the same state in an unknown number of steps, and not with a fixed periodicity, it will have a unique limiting distribution (see Meyn and Tweedie 1993 for proof and explanation)\footnote{If the space of states is continuous, also the so-called ``positive recurrance'' property is needed, meaning that, starting from any state, we can always go back in a finite number of steps.}.

A Markov chain is said \textit{(time) reversible} if the transition kernel satisfies the so-called \textbf{\textit{detailed balance}} condition
\begin{equation}
    \pi(x) \mathcal{T}(x,y) = \pi(y) \mathcal{T}(y,x), 
    \label{eq:det_b_mc}
\end{equation}
which basically means that the probability of being in a state $x$ and going to a state $y$, proposed with the transition kernel $\mathcal{T}(x,y)$, must be the same as being in a state $y$ and going back to the state $x$.
Any Markov chain with a transition kernel $\mathcal{T}$ that satisfies Eq.~\eqref{eq:det_b_mc} will have a stationary distribution $\pi$.
This can be shown by reading the stationary distribution condition simply as ``if we are in a state $x$ described by the limiting distribution $\pi$, and we go to a new state $y$ through the transition kernel $\mathcal{T}(x,y)$, also the new state $y$ will be described by the limiting distribution $\pi$''.

The probability of being at a state $y$ can then be computed as the probability of being in $x$ and going to $y$, for all possible states $x$, i.e.,
\begin{equation}
    p(y) = \int dx \pi(x) \mathcal{T}(x,y) 
    = \int dx \pi(y) \mathcal{T}(y,x) 
    = \pi(y) \int dx \mathcal{T}(y,x) = \pi(y),
\end{equation}
where, in the last step, we used the fact that the transition kernel is normalized $\int \mathcal{T}(y,x) dx =1$. 
Therefore, the detailed balance condition ensures that the probability of the new state $y$ is also described by the limiting distribution $\pi$.

In MCMC samplers, the goal is to build Markov chains whose stationary distribution corresponds to the distribution we want to reconstruct. 
In Bayesian inference, the desired stationary distribution is the posterior.

There exist different MCMC algorithms; in the following, we will discuss the Metropolis-Hastings one~\cite{10.1093/biomet/57.1.97, 1953JChPh..21.1087M}, which can be then easily generalized to the reversible jump case.

\subsection{Metropolis-Hastings algorithm}
\label{ssec:mh}

The Metropolis-Hastings algorithm for MCMC sampling can be summarized as follows. 
Suppose we are in a state $x_i$ in the Markov Chain, to move to the next state $x_{i+1}$ one needs to:
\begin{enumerate}
    \item Propose the new state $x_{i+1}$, generating it from the proposal distribution $q(x_{i+1}|x_i)$, which describes the probability of $x_{i+1}$ given $x_i$ (and only $x_i$, since Markov chains depend only on the previous point).
    \item Decide whether to accept the new state or not, based on the \emph{acceptance probability} $\alpha$. To decide whether the proposed point is accepted, a random value $r$ between $[0,1]$ is drawn: if $\alpha \ge r$, the proposed $x_{i+1}$ is accepted and saved as the next state in the chain, otherwise it is rejected and $x_i$ is saved again as the chain state.
\end{enumerate}

The acceptance ratio is computed as 
\begin{equation}
    \alpha = \min \left\{ 1,  \frac{\pi(\theta_{i+1}) q(\theta_{i+1}|\theta_i)}{\pi(\theta_{i}) q(\theta_{i}|\theta_{i+1})}\right\}.
    \label{eq:acc_mh}
\end{equation}
The expression for the acceptance probability is not unique, but Ref.~\cite{10.1093/biomet/60.3.607} showed how this form optimizes efficiency.

The acceptance probability in Eq.~\eqref{eq:acc_mh} was derived by Hastings in Ref.~\cite{hastings} in the following way:
assume that the chain is in a state $x$ and wants to go to a state $y$ in the same space. 
Based on the Metropolis-Hastings algorithm, the transition kernel can be written as $\mathcal{T}(x,y) = q(y|x) \alpha(x,y)$, where we explicitly wrote the two probabilities of proposing a new point, $q(x|y)$, and accepting it, $\alpha(x,y)$.
The \textit{detailed balance} condition (Eq.~\eqref{eq:det_b_mc}) now reads
\begin{equation}
    \pi(x) q(y|x) \alpha(x,y) = \pi(y) q(x|y) \alpha(y,x).
    \label{eq:det_bal_acc}
\end{equation}

Hastings proposed to write the acceptance probability as
\begin{equation}
    \alpha(x,y) = \frac{s(x,y)}{1 + \frac{\pi(x)q(y|x)}{\pi(y)q(x|y)}},
\end{equation}
with the only condition that $s$ must be symmetric, i.e., $s(x,y) = s(y,x)$. With this expression for the acceptance ratio, Eq.~\eqref{eq:det_bal_acc} can be rewritten as
\begin{align}
    &\pi(x) q(y|x) \frac{s(x,y)}{\frac{\pi(y)q(x|y) + \pi(x)q(y|x)}{\pi(y)q(x|y)}} = \pi(y) q(x|y) \frac{s(y,x)}{\frac{\pi(x)q(y|x) + \pi(y)q(x|y)}{\pi(x)q(y|x)}} \\
    &\frac{\pi(x)q(y|x)s(x,y)\pi(y)q(x|y)}{\pi(y)q(x|y) + \pi(x)q(y|x)} = \frac{\pi(y)q(x|y)s(y,x)\pi(x)q(y|x)}{\pi(x)q(y|x) + \pi(y)q(x|y)}.
\end{align}
Therefore, provided that $s(x,y)$ is symmetric, this form of the acceptance probability ensures that detailed balance is satisfied.

An easy choice for $s(x,y)$ is
\begin{equation}
    s(x,y) = \begin{cases}
        1 + \frac{\pi(x) q(y|x)}{\pi(y) q(x|y)} & \mbox{if} \;\; \frac{\pi(y) q(x|y)}{\pi(x) q(y|x)} \ge 1 \\
        1 + \frac{\pi(y) q(x|y)}{\pi(x) q(y|x)} & \mbox{if} \;\; \frac{\pi(y) q(x|y)}{\pi(x) q(y|x)} < 1.
    \end{cases}
\end{equation}
Therefore, the acceptance ratio becomes
\begin{equation}
    \alpha(x,y) = \begin{cases}
        1  & \mbox{if} \;\; \frac{\pi(y) q(x|y)}{\pi(x) q(y|x)} \ge 1 \\
        \frac{1 + \frac{\pi(y) q(x|y)}{\pi(x) q(y|x)}}{1 + \frac{\pi(x) q(y|x)}{\pi(y) q(x|y)}} & \mbox{if} \;\; \frac{\pi(y) q(x|y)}{\pi(x) q(y|x)} < 1,
    \end{cases}
\end{equation}
or
\begin{equation}
 \alpha(x,y) = \min \left\{  1, \frac{\pi(y) q(x|y)}{\pi(x) q(y|x)} \right\}.
\end{equation}

This expression can be further simplified if the chosen proposal function is symmetric, i.e., if $q(x|y) = q(y|x)$. In this case, the acceptance probability simply becomes
\begin{equation}
     \alpha(x,y) = \min \left\{  1, \frac{\pi(y)}{\pi(x)} \right\}.
\end{equation}

\subsection{Generalizing Metropolis-Hastings}

To facilitate the extension to the trans-dimensional case, we first consider the Metropolis-Hastings MCMC in a more general way.
We now consider two states $x$ and $x'$, belonging to the sets $A$ and $A'$, respectively, which for now we assume are subsets (more precisely Borel sets) of a larger parameter space, e.g., $A, A' \subseteq \mathbb{R}^n$.

We impose detailed balance as the condition that the probability of a chain to be in a state in set $A$ and going to a state in the set $A'$, must be the same as being in a state in $A'$ and going to a state in the set $A$. 
This can be written as
\begin{equation}
    \int_A \pi(x) P(x,A') dx = \int_{A'} \pi(x') P(x',A) dx'.
    \label{eq:det_bal_general}
\end{equation}
The transition kernel is now $P(x,A')$ (and similarly $P(x',A)$), which describes the probability of being in a state $x$ and ending up in the set $A'$.
This can happen in two ways:
\begin{enumerate}
    \item The chain is a state $x \in A$, it proposes a move to $x' \in A'$, and the move is accepted
    \item The chain is in a state $x \in A'$, it proposes a move anywhere but the move is not accepted, therefore the chain remains in the set $A'$. 
\end{enumerate}
This can be written as
\begin{equation}
    P(x,A') = \underbrace{\int_{A'} q(x' | x) \alpha(x,x') dx'}_{\rm case \; 1} 
    + \underbrace{\left[ \int q(x'|x) [1 - \alpha(x,x')] dx' \right] \cdot I_{\left\{ x \in A' \right\}}}_{\rm case \; 2},
\end{equation}
where $[1 - \alpha(x,x')]$ denotes the probability of \emph{not} accepting the move to $x'$, and $I_{\left\{ x \in A' \right\}}$ is the indicator function, i.e.,
\begin{equation}
    I_{\left\{ x \in A' \right\}} = \begin{cases}
        1 & \mbox{if} \;\; x \in A' \\
        0 & \mbox{if} \;\; x \notin A'.
    \end{cases}
\end{equation}

Explicitly writing these two contributions in the detailed balance Eq.~\eqref{eq:det_bal_general}, we obtain
    \begin{equation}
        \int_A dx \, \pi(x) \left[ \int_{A'} q(x'|x) \alpha(x,x') dx' + r(x) \cdot I_{\left\{ x \in A' \right\}} \right] = \int_{A'} dx' \, \pi(x') \left[ \int_{A} q(x|x') \alpha(x',x) dx + r(x') \cdot I_{\left\{ x' \in A \right\}} \right],
    \end{equation}
where we defined
\begin{equation}
    r(x) = \int q(x'|x) [1 - \alpha(x,x')] dx'
\end{equation}
and 
\begin{equation}
    r(x') = \int q(x|x') [1 - \alpha(x',x)] dx.
\end{equation}
Using
\begin{equation}
    \int_A dx \cdot I_{\left\{ x \in A' \right\}} = \int_{A \cap A'} dx,
\end{equation}
the equation above becomes
\begin{equation}
        \int_A \int_{A'} dx dx'\, \pi(x)  q(x'|x) \alpha(x,x')  + \int_{A \cap A' } dx \, \pi(x) r(x) = \int_{A'} \int_A dx' dx \, \pi(x') q(x|x') \alpha(x',x)  + \int_{A \cap A' } dx' \, \pi(x') r(x').
\end{equation}

The second terms on the right- and left-hand side are the same, therefore it reduces to
    \begin{equation}
        \int_A \int_{A'} dx dx' \, \pi(x)  q(x'|x) \alpha(x,x')  = \int_{A'} \int_A dx'  dx \, \pi(x') q(x|x') \alpha(x',x),\
        \label{eq:int_det_bal}
    \end{equation}
which is called the \textit{integrated detailed balance} condition.
One can show that if Eq.~\eqref{eq:int_det_bal} holds for all subsets $A,A' \subseteq \mathbb{R}^n$, then the ``regular" detailed balance (e.g., Eq.~\eqref{eq:det_b_mc}) also holds.

\subsection{The trans-dimensional case}
\label{ssec:tdim_mcmc}

So far we have assumed that the Markov chains always move between spaces with the same dimensionality. However, especially in the case where we want to compare different models, the states involved might belong to spaces with different dimensions. 
A ``solution" to this issue, proposed by Green in Ref.~\cite{rjmcmc}, consists in introducing so-called \emph{auxiliary-variables} to ensure that, in the end, we always move between spaces with the same dimensionality.

Suppose we want to move between a set with dimension $n_k$ and another one with dimensions $n_{k'}$, i.e., $A \subseteq \mathbb{R} ^{n_k}$ and $A' \subseteq \mathbb{R}^{n_{k'}}$. 
When the chain is in a state $x \in A$, we generate a random variable $\vec{u}$ with dimension $r$ from a distribution $g(\vec{u})$, and, similarly, a random variable $\vec{u}'$ with dimension $r'$ from a distribution $g'(\vec{u}')$ for the chain that is at the state $x'$ in $A'$. We define a mapping $h$ between the states and auxiliary variables of the two sets, such that
\begin{equation}
    (x', \vec{u}') = h(x, \vec{u}) \hspace{1cm}  (x, \vec{u}) = h^{-1}(x', \vec{u}'),
    \label{eq:det_map}
\end{equation}
where $h^{-1}$ is the inverse function of $h$.

To ensure that the detailed balance is respected, the only condition upon the deterministic mapping $h$ is to be a diffeomorphism, which in turn requires \emph{dimension-matching}, i.e.,
\begin{equation}
    n_k + r = n_{k'} + r'.
\end{equation}

Now let us look at the detailed balance condition in Eq.~\eqref{eq:int_det_bal}. When working with auxiliary variables, the probability of proposing a new state reduces to the probability of drawing the specific random variable involved in the proposal, since the jump is then dictated by the deterministic transformation (Eq.~\eqref{eq:det_map}). If the chain is in the current state $x$, the new point $x'$ is proposed through the function $h(x,\vec{u})$.
Since $h$ is a deterministic mapping, it in turn depends on the drawn value for the random auxiliary variable $\vec{u}$, thus the probability of proposing the new point $x'$ corresponds to the probability of drawing a specific value $\vec{u}$, i.e., it is given by $g(\vec{u})$. Therefore, when working with maps between spaces, Eq.~\eqref{eq:int_det_bal} can be rewritten as
\begin{equation}
    \int_A \int_{A'} \pi(x) g(\vec{u}) \alpha(x,x') \, dx d\vec{u} = \int_{A'} \int_A  \pi(x') g'(\vec{u}') \alpha(x',x) dx' d\vec{u}',
    \label{eq:det_bal_h}
\end{equation}
where we remind that $\pi(x)$ is the probability of being in state $x$, $g(\vec{u})$ is the probability of drawing the random variable $\vec{u}$, and $\alpha(x,x')$ represents the probability to accept the jump from state $x$ to state $x'$, and similarly for the right-hand-side of the equation.

Since we demand that the mapping $h$ is a diffeomorphism, $h$ and its inverse $h^{-1}$ are both differentiable, we can perform a change of variables in the right-hand-side integral in Eq.~\eqref{eq:det_bal_h} as $dx' d\vec{u}' = \frac{\partial h(x,\vec{u})}{\partial(x,\vec{u})} dx d\vec{u}$, obtaining
\begin{equation}
    \int_A \int_{A'} \pi(x) g(\vec{u}) \alpha(x,x') \, dx d\vec{u} = \int_{A'} \int_A  \pi(x') g'(\vec{u}') \alpha(x',x) \left| \frac{\partial h(x,\vec{u})}{\partial(x,\vec{u})} \right| dx d\vec{u},
\end{equation}
where the term $ \left| \frac{\partial h(x,\vec{u})}{\partial(x,\vec{u})} \right|$ is the Jacobian arising from the change of variables between $(x, \vec{u})$ and $(x', \vec{u}')$.
This implies
\begin{equation}
    \pi(x) g(\vec{u}) \alpha(x,x') = \pi(x') g'(\vec{u}') \alpha(x',x) \left| \frac{\partial h(x,\vec{u})}{\partial(x,\vec{u})} \right|,
    \label{eq:det_bal_acc_rjmcmc}
\end{equation}
and, repeating the derivation in Sec.~\ref{ssec:mh}, we find that the acceptance probability now reads
\begin{equation}
 \alpha(x,x') = \min \left\{  1, \frac{\pi(x') g'(\vec{u}')}{\pi(x) g(\vec{u})} \left| \frac{\partial h(x,\vec{u})}{\partial(x,\vec{u})} \right| \right\}.
 \label{eq:acc_rj}
\end{equation}


\subsection{Reversible Jump MCMC}
\label{ssec:rjmcmc}

As discussed earlier, when using RJMCMC algorithms in Bayesian inference to compare models or hypothesis is to sample the joint posterior probability $p(\vec{\theta}_k, k | d)$ of the parameters \emph{and} the model describing the data. Therefore, the chains move between states given by $x = (\vec{\theta}_k, k)$, with $k$ being the label denoting the model and $\vec{\theta}_k$ the parameters vector for that specific model. At each step the sampler has to perform two kinds of moves:
\begin{enumerate}
    \item \textbf{Within model move}: Within a fixed model $k$, update the model's parameters $\vec{\theta}_k$ with a standard MCMC procedure (e.g., the Metropolis Hastings algorithm described above).
    \item \textbf{Between-models move}: Propose a jump to a different model, thus updating $k$ and $\vec{\theta}_k$ simultaneously.
\end{enumerate}

For the \textit{within-model moves}, a ``standard" Metropolis-Hastings algorithm can be employed. In this move we want to go from a state $x=(\vec{{\theta}_k}, k)$ to a state $(\vec{\theta}'_k, k)$. Since the model $k$ is fixed, we can drop its identifying label in the states. Writing explicitly the stationary distribution $\pi$ that we want to reconstruct as the joint posterior $p(\vec{\theta}_k, k | d)$, we find that the acceptance ratio in Eq.~\eqref{eq:acc_mh} becomes
\begin{equation}
  \alpha(\vec{\theta}_k, \vec{\theta}'_k) = \min \left\{ 1, \frac{p(\vec{\theta}'_k | d) q(\vec{\theta}_k|\vec{\theta}'_k)}{p(\vec{\theta}_k | d) q(\vec{\theta}'_k|\vec{\theta}_k)}\right\} = \min \left\{ 1, \frac{\mathcal{L}(d | \vec{\theta}'_k, k) p(\vec{\theta}'_k|k) q(\vec{\theta}_k|\vec{\theta}'_k)}{\mathcal{L}(d | \vec{\theta}_k, k) p(\vec{\theta}_k|k) q(\vec{\theta}'_k|\vec{\theta}_k)} \right\},
\end{equation}
where in the second step we applied Bayes' theorem (see Eq.~\eqref{eq:bayes_mm}), and $\mathcal{L}$ denotes the likelihood term. 
For the prior, we used the fact that the joint model-parameters prior can be factorized in $p(k,\vec{\theta}_k) = p(k) p(\vec{\theta}_k|k)$, and the $p(k)$ terms cancel out because the chain is moving within the same model.
Since the evidence is a normalization constant in the posterior probabilities, it is the same across all states, therefore it cancels out in the acceptance probability, and we need to compute only the likelihood and the prior of each state.

In the \textit{between-model move}, instead, the acceptance ratio derived for the general and trans-dimensional case in Eq.~\eqref{eq:acc_rj} must be used. This time, the chain is in a state $x = (\vec{\theta}_k, k)$ and wants to move to $x' = (\vec{\theta}'_{k'}, k')$. Again writing explicitly the stationary distribution as the joint posterior $p(\vec{\theta}_k, k | d)$, we find
\begin{equation}
    \alpha(x,x') = \min \left\{  1, \frac{\mathcal{L}(d | \vec{\theta}'_{k'}, k') p(k', \vec{\theta}'_{k'}) }{ \mathcal{L}(d | \vec{\theta}_{k}, k) p(k, \vec{\theta}_{k})} \frac{g'(\vec{u}')}{g(\vec{u})} \left| \frac{\partial h(\vec{\theta}_k,\vec{u})}{\partial(\vec{\theta}_k,\vec{u})} \right| \right\} = \min \left\{  1, \frac{\mathcal{L}(d | \vec{\theta}'_{k'}, k') p(k') p(\vec{\theta}'_{k'} | k') }{ \mathcal{L}(d | \vec{\theta}_{k}, k) p(k) p(\vec{\theta}_{k} | k)} \frac{g'(\vec{u}')}{g(\vec{u})} \left| \frac{\partial h(\vec{\theta_k},\vec{u})}{\partial(\vec{\theta}_k,\vec{u})} \right| \right\}, 
\end{equation}
where the evidence cancels out. We also used the fact that the state prior can be written as $p(k, \vec{\theta}_k) = p(k) p(\vec{\theta}_k | k)$, however, this time the term describing the model prior does not cancel out since different models can have different prior probabilities.

\section{RJMCMC details for t-roo}
\label{sec:app_rjmcmc_troo}

\subsection{Full proposal derivation and inverse moves}
\label{sec:app_proposals}

In the following, we show how the move proposals in \Cref{eq:bbh_to_nsbh,eq:bns_to_bbh,eq:nsbh_to_bns} in Sec.~\ref{sec:proposals} are derived, and how we obtain the proposals for the inverse move.

Let us start from the proposal for the tidal parameters and look at the case BNS$\rightarrow$BBH as an example.
As discussed in the main text, we update the tidal parameter through a stretch move according to Eq.~\eqref{eq:bns_to_bbh}.
If now we want to go back to the BNS case, we write the proposal for the new $\Lambda_1'$ in the BNS model as
\begin{equation}
    \Lambda'_{1,\textsc{bns}} = \Lambda_{1,j} + u_1 \left( \Lambda_{1,\textsc{bbh}} - \Lambda_{1,j}\right).
    \label{eq:lam_bbh_inv}
\end{equation}
For the transformation to be reversible, we want that this proposal allows us to go back to the ``original'' point in the BNS model, i.e., to the $\Lambda_{1, \textsc{bns}}$ from which we proposed the move to $\Lambda_{1, \textsc{bbh}}$ in Eq.~\eqref{eq:bbh_to_nsbh}. 
Substituting the $\Lambda_{1, \textsc{bbh}}$ proposal in Eq.~\eqref{eq:bns_to_bbh} into Eq.~\eqref{eq:lam_bbh_inv} leads to
\begin{align}
    \Lambda'_{1,\textsc{bns}} &= \Lambda_{1,j} + u_1 \left( \Lambda_{1,j} + w_1\left( \Lambda_{1, \textsc{bns}} - \Lambda_{1,j} \right)  -\Lambda_{1,j} \right) \\
    &= \Lambda_{1,j} - u_1 w_1 \Lambda_{1,j} + u_1 w_1 \Lambda_{1,\textsc{bns}} \\
    &= (1 - u_1 w_1) \Lambda_{1,j} + u_1 w_1 \Lambda_{1,\textsc{bns}},
\end{align}
This shows that the condition for this move to be reversible is
\begin{equation}
    u_1 w_1 = 1.
\end{equation}
Similarly, one can show that the reversibility condition for the $\Lambda_2$ proposal reads
\begin{equation}
    u_2 w_2 = 1.
\end{equation}
Therefore the move between auxiliary variables is
\begin{equation}
    \mbox{BNS $\rightarrow$ BBH:} \; \; u_1 = \frac{1}{w_1}, \; u_2\ = \frac{1}{w_2} \hspace{0.5cm} \mbox{and} \hspace{0.5cm} \mbox{BBH $\rightarrow$ BNS:} \; \; w_1 = \frac{1}{u_1}, \; w_2 = \frac{1}{u_2}.
\end{equation}
In the case of moves between an NSBH and a BBH model, both \lonebb{} and \lonenb{} are \textit{pseudo-}parameters.
Hence, we do not propose a stretch move in this case and instead just keep the same values when jumping to the other model.
This also implies that the auxiliary variables $u_1, v_1$ are not needed in this proposal\footnote{We remind that we have auxiliary variables $(u_1, u_2)$, $(v_1,v_2)$, and $(w_1, w_2)$ in the BBH, NSBH, and BNS space, respectively.}.

For the chirp mass and mass ratio proposals, the idea, outlined in Sec.~\ref{subsec:between_models_move},is to move along a line with slope determined by the likelihood center-of-mass samples in the preliminary sampling, translating this line in the $\mathcal{M}_c - \tilde{\Lambda}$ or $q-\tilde{\Lambda}$ space to make it pass through the current sample point. Following the slope definition in Eq.~\eqref{eq:slope_mc}, the line equation, in the BBH$\rightarrow$BNS case, reads
\begin{equation}
    (\mathcal{M}_{c,\textsc{bbh}} - \mathcal{M}_{c,\textsc{bns}}) = s_{\mathcal{M}_{c,(\textsc{bbh},\textsc{bns})}} \cdot (\bar{\tilde{\Lambda}}_\textsc{bbh} - \tilde{\Lambda}_\textsc{bns}),
\end{equation}
where the bar on $\tilde{\Lambda}$ denotes that $\tilde{\Lambda}$ is computed assuming, for the \textit{pseudo}-parameters, the physical values $\Lambda_{1, \textsc{bbh}} = \Lambda_{2, \textsc{bbh}} = \Lambda_{1,\textsc{nsbh}} = 0$, which therefore imply $\bar{\tilde{\Lambda}}_{\textsc{bbh}}=0$.
Therefore, the chirp-mass proposal in Eq.~\eqref{eq:bns_to_bbh} becomes
\begin{align}
    \mathcal{M}_{c, \textsc{bbh}} &= \mathcal{M}_{c, \textsc{bns}} + s_{\mathcal{M}_c \left( \textsc{bbh}, \textsc{bns} \right)}\left( \bar{\tilde{\Lambda}}_\textsc{bbh} - \tilde{\Lambda}_{\textsc{bns}}\right) \\
    &= \mathcal{M}_{c, \textsc{bns}} - s_{\mathcal{M}_c \left( \textsc{bbh}, \textsc{bns} \right)} \tilde{\Lambda}_{\textsc{bns}},
    \label{eq:mc_bbh_for}
\end{align}
and similarly we can derive the proposal for the mass ratio.

Since the mapping in RJMCMC must be invertible, the proposals for the moves in the opposite direction, i.e., BBH$\rightarrow$BNS, can be found by simply inverting this mapping. 

In the specific case of the chirp mass proposal when going from the BBH to the BNS model, we have to use
\begin{equation}
    \mathcal{M}'_{c,\textsc{bns}} = \mathcal{M}_{c,\textsc{bbh}} + s_{\mathcal{M}_c \left( \textsc{bbh}, \textsc{bns} \right)} \tilde{\Lambda}'_{\textsc{bns}},
    \label{eq:mc_bns_back}
\end{equation}
where the prime denotes the new point proposed in the BNS space. 
For the transformation to be reversible, we need to ensure that we can go back to the starting point in the BNS space, i.e., that we can have $ \mathcal{M}'_{c,\textsc{bns}} =  \mathcal{M}_{c,\textsc{bns}}$.
Plugging Eq.~\eqref{eq:mc_bbh_for} into Eq.~\eqref{eq:mc_bns_back}
\begin{align}
    \mathcal{M}'_{c,\textsc{bns}} &= \mathcal{M}_{c, \textsc{bns}} - s_{\mathcal{M}_c \left( \textsc{bbh}, \textsc{bns} \right)} \tilde{\Lambda}_{\textsc{bns}} + s_{\mathcal{M}_c \left( \textsc{bbh}, \textsc{bns} \right)} \tilde{\Lambda}'_{\textsc{bns}} \\
    & = \mathcal{M}_{c, \textsc{bns}} - s_{\mathcal{M}_c \left( \textsc{bbh}, \textsc{bns} \right)} \left( \tilde{\Lambda}_{\textsc{bns}} - \tilde{\Lambda}'_{\textsc{bns}} \right).
\end{align}
This shows that the transformation is reversible when $\tilde{\Lambda}_{\textsc{bns}} = \tilde{\Lambda}'_{\textsc{bns}}$, which in turn is ensured by the conditions $u_1w_1 = 1$ and $u_2 w_2 = 1$ derived above.

In the same way one can derive the full proposals for the moves between BBH$\rightarrow$NSBH and NSBH$\rightarrow$BNS (see \Cref{eq:bbh_to_nsbh,eq:nsbh_to_bns}, respectively), and their inverse. 
For completeness, we list the the inverse mappings here

\begin{align}
\textbf{NSBH to BBH} \qquad & \begin{cases}
        \vec{\Theta}_{\textsc{bbh}, i+1} = \vec{\Theta}_{\textsc{nsbh}, i} \\
        \Lambda_{1, \textsc{bbh}, i+1} = \Lambda_{1, \textsc{nsbh}, i} \\
        \Lambda_{2, \textsc{bbh}, i+1} = \Lambda_{2,j, i} + v_2 \left( \Lambda_{2, \textsc{nsbh}, i} - \Lambda_{2,j, i} \right) \\
        \mathcal{M}_{c, \textsc{bbh}, i+1} = \mathcal{M}_{c, \textsc{nsbh}, i} -  s_{\mathcal{M}_c (\textsc{nsbh}, \textsc{bbh})} \cdot \bar{\tilde{\Lambda}}_{\textsc{nsbh}, i} \\
        q_{\textsc{bbh}, i+1} = q_{\textsc{bbh}, i} - s_{q (\textsc{nsbh}, \textsc{bbh})} \cdot \bar{\tilde{\Lambda}}_{\textsc{nsbh}, i} \\
        u_2 = 1/v_2 
        \end{cases} \label{eq:nsbh_to_bbh}  \\[3ex]
\textbf{BBH to BNS} \qquad &    \begin{cases}
        \vec{\Theta}_{\textsc{bns}, i+1} = \vec{\Theta}_{\textsc{bbh}, i} \\
        \Lambda_{1, \textsc{bns}, i+1} = \Lambda_{1,j, i} + u_{1} \left( \Lambda_{1, \textsc{bbh}, i} - \Lambda_{1,j, i} \right) \\
        \Lambda_{2, \textsc{bns}, i+1} = \Lambda_{2,j, i} + u_2 \left( \Lambda_{1, \textsc{bbh}, i} - \Lambda_{2,j, i} \right) \\
        \mathcal{M}_{c, \textsc{bns}, i+1} = \mathcal{M}_{c, \textsc{bbh}, i} + s_{\mathcal{M}_c (\textsc{bbh}, \textsc{bns})} \tilde{\Lambda}_{\textsc{bns}, i+1} \\
        q_{\textsc{bns}, i+1} = q_{\textsc{bbh}, i} + s_{q (\textsc{bbh, \textsc{bns})}} \tilde{\Lambda}_{\textsc{bns}, i+1} \\
        w_1 = 1 /u_1 \\
        w_2 = 1/u_2 
        \end{cases} \label{eq:bbh_to_bns} \\[3ex]
\textbf{BNS to NSBH}\qquad&\begin{cases}
        \vec{\Theta}_{\textsc{nsbh}, i+1} = \vec{\Theta}_{\textsc{bns},i} \\
        \Lambda_{1, \textsc{nsbh}, i+1} = \Lambda_{1,j,i} + w_1 \left( \Lambda_{1, \textsc{bns}, i} - \Lambda_{1,j, i} \right) \\
        \Lambda_{2, \textsc{nsbh}, i+1} = \Lambda_{2,j,i} + w_2 \left( \Lambda_{1, \textsc{bns}, i} - \Lambda_{2,j, i} \right) \\
        \begin{aligned}
        \mathcal{M}_{c, \textsc{nsbh}, i+1} = &\mathcal{M}_{c, \textsc{bns}, i} + s_{\mathcal{M}_c (\textsc{bns}, \textsc{nsbh})}  \left( \bar{\tilde{\Lambda}}_{\textsc{nsbh}, i+1} - {\tilde{\Lambda}}_{\textsc{bns}, i} \right) \end{aligned} \\
        q_{\textsc{bns}, i+1} = q_{\textsc{nsbh}, i} + s_{q (\textsc{bns}, \textsc{nsbh})}  \left( \bar{\tilde{\Lambda}}_{\textsc{nsbh}, i+1} - {\tilde{\Lambda}}_{\textsc{bns}, i} \right) \\
        v_1 = 1 / w_1 \\
        v_2 = 1/w_2
        \end{cases} \label{eq:bns_to_nsbh} \ .
\end{align}

\subsection{Jacobians}
\label{app:jacobians}

As discussed in Sec.~\ref{ssec:tdim_mcmc}, it is essential to include the Jacobian in the acceptance-ratio computation for the between-model move. In the following we compute the Jacobian specifically for the move BNS$\rightarrow$BBH, i.e., for the mapping in Eq.~\eqref{eq:bns_to_bbh}. The Jacobians for all the other transformations can be computed with the same procedure.

The Jacobian for the transformation between the BNS and BBH space, i.e., from $(\vec{\Theta}_\textsc{bns}, \Lambda_{1, \textsc{bns}}, \Lambda_{2, \textsc{bns}}, \mathcal{M}_{c, \textsc{bns}}, q_\textsc{bns}, w_1, w_2)$ to $(\vec{\Theta}_\textsc{bbh}, \Lambda_{1, \textsc{bbh}}, \Lambda_{2, \textsc{bbh}}, \mathcal{M}_{c, \textsc{bbh}}, q_\textsc{bbh}, u_1, u_2)$, is given by
\begin{equation}
    \mathbb{J}_{\textsc{bns,bbh}} = \begin{vmatrix}
        \frac{\partial \vec{\Theta}_\textsc{bbh}}{\partial \vec{\Theta}_\textsc{bns}} & \frac{\partial \vec{\Theta}_\textsc{bbh}}{\partial \Lambda_{1, \textsc{bns}}} & \frac{\partial \vec{\Theta}_\textsc{bbh}}{\partial \Lambda_{2, \textsc{bns}}} & \frac{\partial \vec{\Theta}_\textsc{bbh}}{\partial \mathcal{M}_{c, \textsc{bns}}} & \frac{\partial \vec{\Theta}_\textsc{bbh}}{\partial q_\textsc{bns}} & \frac{\partial \vec{\Theta}_\textsc{bbh}}{\partial w_1} & \frac{\partial \vec{\Theta}_\textsc{bbh}}{\partial w_2} \\[1.5ex]

        \frac{\partial \Lambda_{1, \textsc{bbh}}}{\partial \vec{\Theta}_\textsc{bns}} & \frac{\partial \Lambda_{1, \textsc{bbh}}}{\partial \Lambda_{1, \textsc{bns}}} & \frac{\partial \Lambda_{1, \textsc{bbh}}}{\partial \Lambda_{2, \textsc{bns}}} & \frac{\partial \Lambda_{1, \textsc{bbh}}}{\partial \mathcal{M}_{c, \textsc{bns}}} & \frac{\partial \Lambda_{1, \textsc{bbh}}}{\partial q_\textsc{bns}} & \frac{\partial \Lambda_{1, \textsc{bbh}}}{\partial w_1} & \frac{\partial \Lambda_{1, \textsc{bbh}}}{\partial w_2} \\[1.5ex]

        \frac{\partial \Lambda_{2, \textsc{bbh}}}{\partial \vec{\Theta}_\textsc{bns}} & \frac{\partial \Lambda_{2, \textsc{bbh}}}{\partial \Lambda_{1, \textsc{bns}}} & \frac{\partial \Lambda_{2, \textsc{bbh}}}{\partial \Lambda_{2, \textsc{bns}}} & \frac{\partial \Lambda_{2, \textsc{bbh}}}{\partial \mathcal{M}_{c, \textsc{bns}}} & \frac{\partial \Lambda_{2, \textsc{bbh}}}{\partial q_\textsc{bns}} & \frac{\partial \Lambda_{2, \textsc{bbh}}}{\partial w_1}  & \frac{\partial \Lambda_{2, \textsc{bbh}}}{\partial w_2}  \\[1.5ex]

        \frac{\partial \mathcal{M}_{c, \textsc{bbh}}}{\partial \vec{\Theta}_\textsc{bns}} & \frac{\partial \mathcal{M}_{c, \textsc{bbh}}}{\partial \Lambda_{1, \textsc{bns}}} & \frac{\partial \mathcal{M}_{c, \textsc{bbh}}}{\partial \Lambda_{2, \textsc{bns}}} & \frac{\partial \mathcal{M}_{c, \textsc{bbh}}}{\partial \mathcal{M}_{c, \textsc{bns}}} & \frac{\partial \mathcal{M}_{c, \textsc{bbh}}}{\partial q_\textsc{bns}} & \frac{\partial \mathcal{M}_{c, \textsc{bbh}}}{\partial w_1} & \frac{\partial \mathcal{M}_{c, \textsc{bbh}}}{\partial w_2} \\[1.5ex]

        \frac{\partial q_\textsc{bbh}}{\partial \vec{\Theta}_\textsc{bns}} & \frac{\partial q_\textsc{bbh}}{\partial \Lambda_{1, \textsc{bns}}} & \frac{\partial q_\textsc{bbh}}{\partial \Lambda_{2, \textsc{bns}}} & \frac{\partial q_\textsc{bbh}}{\partial \mathcal{M}_{c, \textsc{bns}}} & \frac{\partial q_\textsc{bbh}}{\partial q_\textsc{bns}} & \frac{\partial q_\textsc{bbh}}{\partial w_1} & \frac{\partial q_\textsc{bbh}}{\partial w_2} \\[1.5ex]

        \frac{\partial u_1}{\partial \vec{\Theta}_\textsc{bns}} & \frac{\partial u_1}{\partial \Lambda_{1, \textsc{bns}}} & \frac{\partial u_1}{\partial \Lambda_{2, \textsc{bns}}} & \frac{\partial u_1}{\partial \mathcal{M}_{c, \textsc{bns}}} & \frac{\partial u_1}{\partial q_\textsc{bns}} & \frac{\partial u_1}{\partial w_1} & \frac{\partial u_1}{\partial w_2} \\[1.5ex]

        \frac{\partial u_2}{\partial \vec{\Theta}_\textsc{bns}} & \frac{\partial u_2}{\partial \Lambda_{1, \textsc{bns}}} & \frac{\partial u_2}{\partial \Lambda_{2, \textsc{bns}}} & \frac{\partial u_2}{\partial \mathcal{M}_{c, \textsc{bns}}} & \frac{\partial u_2}{\partial q_\textsc{bns}} & \frac{\partial u_2}{\partial w_1} & \frac{\partial u_2}{\partial w_2}. \\[1.5ex]
        
    \end{vmatrix}
    \label{eq:jac_bns_to_bbh}
\end{equation}

Looking at the mapping in Eq.~\eqref{eq:bns_to_bbh} and computing all the derivatives, we find
\begin{align}
\begin{split}
   \mathbb{J}_{\textsc{bns,bbh}} &= \begin{vmatrix}
        \mathds{1} & 0 & 0 & 0 & 0 & 0 & 0 \\
        0 & w_1 & 0 & 0 & 0 & \left( \Lambda_{1, \textsc{bns}} - \Lambda_{1,j}\right) & 0 \\
        0 & 0 & w_2 & 0 & 0 & 0 & \left( \Lambda_{2, \textsc{bns}} - \Lambda_{2,j}\right) \\
        0 & -s_{\mathcal{M}_{c, \left( \textsc{bbh, bns}\right)}} \left| \frac{\partial \tilde{\Lambda}_{\textsc{bns}}}{\partial \Lambda_{1,\textsc{bns}}}\right| & -s_{\mathcal{M}_{c, \left( \textsc{bbh, bns}\right)}} \left| \frac{\partial \tilde{\Lambda}_{\textsc{bns}}}{\partial \Lambda_{2,\textsc{bns}}}\right| & 1 & 0 & 0 & 0 \\
        0 & -s_{q_{ \left( \textsc{bbh, bns}\right)}} \left| \frac{\partial \tilde{\Lambda}_{\textsc{bns}}}{\partial \Lambda_{1,\textsc{bns}}}\right| & -s_{q_{\left( \textsc{bbh, bns}\right)}} \left| \frac{\partial \tilde{\Lambda}_{\textsc{bns}}}{\partial \Lambda_{2,\textsc{bns}}}\right| & 0 & 1 & 0 & 0 \\
        0 & 0 & 0 & 0 & 0 & -\frac{1}{w_1^2} & 0 \\
        0 & 0 & 0 & 0 & 0 & 0 & -\frac{1}{w_2^2}
    \end{vmatrix} \\
    &= w_1 w_2 \begin{vmatrix}
        1 & 0 & 0 & 0 \\
        0 & 1 & 0 & 0 \\
        0 & 0 & -\frac{1}{w_1^2} & 0 \\
        0 & 0 & 0 & -\frac{1}{w_2^2}
    \end{vmatrix} = \frac{1}{w_1 w_2} \ .
\end{split}
\end{align}

When we compute the acceptance ratio for this move, i.e., Eq.~\eqref{eq:between_acc}, we need to include this Jacobian and the probability of drawing the auxiliary variables, which are given by $g(z)= 1/\sqrt{z}$ (see Sec.~\ref{sec:proposals}). 
For the auxiliary variables in the BNS we have $g_\textsc{bns}(w_1)=1/\sqrt{w_1}$ and $g_\textsc{bns}(w_2)=1/\sqrt{w_2}$. 
In the BBH space, instead, $g_\textsc{bbh}(u_1)=1/\sqrt{u_1} = \sqrt{w_1}$ and $g_\textsc{bbh}(u_2)=1/\sqrt{u_2} = \sqrt{w_2}$, where we have used the reversibility condition $u_1 w_1 = 1$ and $u_2 w_2 = 1$.
Together with the Jacobian, the factor in the acceptance probability becomes
\begin{align}
    \frac{g_\textsc{bbh}(u_1,u_2)}{g_\textsc{bns}(w_1, w_2)} \mathbb{J}_{\textsc{bns,bbh}} = \frac{\sqrt{w_1 w_2}}{\frac{1}{\sqrt{w_1 w_2}}} \cdot \frac{1}{w_1 w_2} = 1. 
\end{align}
Similarly one can compute the Jacobian for all the other moves, finding in all cases
\begin{equation}
    \frac{g'(x')}{g(x)} \mathbb{J}_{xx'} = 1\ ,
\end{equation}
as shown in Eq.~\eqref{eq:factor}.

\subsection{Symmetry conditions for stretch moves}
\label{ssec:app_symm_stretch}

The stretch move employed in \texttt{t-roo} proposals is described in Sec.~\ref{sec:stretch_move}.
The acceptance ratio for the stretch move shown in Eq.~\eqref{eq:acc_stretch} includes a term to preserve the symmetry condition in the proposal. 
Note that this is different from the detailed balance condition, which includes in its definition the acceptance ratio itself. 
Here, we are interested in enforcing that the kernel to propose a new point given the current one is symmetric. 
This simplifies the calculation of the acceptance probability in the algorithm, but is not a necessary condition such as the detailed balance one, which is instead needed to ensure that the limiting distribution of the chain is stationary.

Assume the chain is in $X_k$ and we use a stretch move to propose a new point $Y$ with
\begin{equation}
    Y = X_j + z (X_k - X_j),
    \label{eq:propY}
\end{equation}
where $X_j$ is the current position of another chain chosen randomly and $z$ is a random variable drawn from a distribution $g(z)$.
Then, from $Y$ we propose a new point $Y'$ with
\begin{equation}
    Y' = X_j + z' (Y - X_j).
    \label{eq:propback}
\end{equation}
To ensure that the proposal is symmetric, i.e., that the probability of going in one direction or the other is the same, we first then need to check whether it is possible that  $Y' = X_k$. 
Combining Eq.~\eqref{eq:propY} and Eq.~\eqref{eq:propback}
\begin{align}
    Y' &= X_j + z' (Y - X_j) \\\
    &= X_j + z' (X_j +z (X_k -X_j) - X_j) \\
    & = X_j + z'z X_k - zz'X_j.
\end{align}
Therefore, in order to have $Y' = X_k$, we need $zz' = 1$.

To find the general symmetry condition for the proposal, we now consider the probability of proposing one point from another one, integrated over all the possible starting points. For the proposals in Eqs.~\eqref{eq:propY}-\eqref{eq:propback},
\begin{align}
  &  \int P({Y}|{X}_k) d{Y} = \int g(z)dz , \\
  & \int P({Y'}|{Y}) d{Y'} = \int g(z') dz',
  \label{eq:prob}
\end{align}
since the probability of a specific point being proposed with a stretch move corresponds to the probability of a certain value for $z$ being chosen.

Stretch moves can be pictured as moving on a sphere with the radius given by the distance to the position of the other randomly selected chain $X_j$, and centered in the origin $X_j$, and. Switching to radial coordinates, in a $n$-dimensional space
\begin{align}
    d{Y} = d \Omega_{n-1} /, dr /, r^{n-1} \; \: \text{with radius} \; r = z | X_k - X_j | ^n \\
    d{Y'} = d \Omega_{n-1} /, dr' /, r'^{n-1} \; \; \text{with radius} \; r' = z' | Y - X_j | ^n .
\end{align}
Changing variables with $dr = |X_k -X_j| dz$ and $dr' = |Y - X_j| dz'$, from Eqs.~\eqref{eq:prob} we find:
\begin{equation}
    P({Y}|{X}_k) = \frac{g(z) dz}{ d\Omega_{n-1} dz |{X}_k - {X}_j| z^{n-1} |{X}_k - {X}_j|^{n-1} }
\end{equation}
and
\begin{equation}
    P({Y'}|{Y}) = \frac{g(z') dz'}{ d\Omega_{n-1} dz' |{Y} - {X}_j| z'^{n-1} |{Y} - {X}_j|^{n-1} }.
\end{equation}

In a symmetric proposal, we want to go back to the starting point, i.e., $P({Y}|{X}_k) = P({Y'}|Y)$. Therefore, we can find the symmetry condition as
\begin{align}
    1 &= \frac{P({Y}|{X}_k)}{P({Y'}|Y)} \\
    &= \frac{g(z)}{g(z')} \frac{z'^{n-1} |{Y} - {X}_j|^n}{z^{n-1} |{X}_k - {X}_j|^n}\\
    &= \frac{g(z)}{g(1/z)} \left( \frac{1}{z}\right) ^{2n -2} \frac{z^n |{X}_k - {X}_j|^n}{|{X}_k - {X}_j|^n} \\
    &= \frac{g(z)}{g(1/z)} z^{2-n},
\end{align}
where in the third line we imposed ${Y'} = {X}_k$, i.e., $z' = 1/z$, and in the end we used $|{Y} - {X}_j|^n = z^n |{X}_k - {X}_j|^n$, from Eq.~\eqref{eq:propY}.

Therefore, the most general condition in $n$ dimensions is:
\begin{equation}
    \frac{g(z)}{g(1/z)} z^{2-n} =1.
\end{equation}
If we consider the one-dimensional case, $n=1$ and
\begin{equation}
    g(z) z = g(1/z).
    \label{eq:symm1d}
\end{equation}

Adding the generic symmetric factor to the acceptance probability in Eq.~\eqref{eq:acc_mh}, allows us to cancel out the specific proposal distributions, i.e.,
\begin{align}
    \alpha &= \min \left\{ 1,  \frac{\pi(\theta_{i+1}) q(\theta_{i+1}|\theta_i)}{\pi(\theta_{i}) q(\theta_{i}|\theta_{i+1})} \frac{g(z)}{g(1/z)} z^{2-n} \frac{g(1/z)}{g(z)}z^{n-2}\right\} \\
    &= \min \left\{ 1,  \frac{\pi(\theta_{i+1})}{\pi(\theta_{i})}  \frac{g(1/z)}{g(z)}z^{n-2}\right\}.
\end{align}
With the specific random variable distribution $g(z)$ in Eq.~\eqref{eq:g_stretch} 
\begin{equation}
    \frac{g(1/z)}{g(z)}z^{n-2} = \frac{1}{\sqrt{1/z}}\frac{1}{1/\sqrt{z}} z^{n-2} = z^{n-1},
\end{equation}
and therefore we recover the expression for the acceptance ratio in Eq.~\eqref{eq:acc_mh}.

\subsection{Priors on pseudo-parameters}
\label{ssec:app_priors}

As explained in Sec.~\ref{sec:proposals}, in \texttt{t-roo} we add the \textit{pseudo-}parameters \lonebb{}, \ltwobb{}, and \lonenb{} to store the information about tidal parameters in models that would not include such parameters. 
In this section, we show that the prior on these \textit{pseudo-}parameters has to (crucially) be included in the acceptance ratio in order to preserve detailed balance.

For simplicity, let us look at the case where we attempt a move between a BBH and NSBH model, with the proposal in Eq.~\eqref{eq:bbh_to_nsbh}. 
The move happens between the spaces $(\vec{\theta}_{\textsc{bbh}}, \Lambda_{1, \textsc{bbh}}, \Lambda_{2, \textsc{bbh}}, u_2)$ for the BBH model and $(\vec{\theta}_{\textsc{nsbh}}, \Lambda_{1, \textsc{nsbh}}, v_2)$ for the NSBH one, with $u_2$ and $v_2$ being the auxiliary variables.
In the detailed balance condition in Eq.~\eqref{eq:det_bal_acc_rjmcmc}
\begin{equation}
    \pi(x) g(\vec{u}) \alpha(x,x') = \pi(x') g'(\vec{u}') \alpha(x', x) \left| \frac{\partial h(x,\vec{u})}{\partial (x,\vec{u})}\right|,
\end{equation}
in this specific case the state $x$ corresponds to $(k={\rm\textsc{bbh}}, \vec{\theta}_k = \vec{\theta}_{\textsc{bbh}})$, while the state $x'$ corresponds to $(k={\rm\textsc{nsbh}}, \vec{\theta}_k = \vec{\theta}_{\textsc{nsbh}})$. 
The stationary distribution $\pi$ corresponds to the posterior probability densities $\pi(x) = p(k=\textsc{bbh},\vec{\theta}_\textsc{bbh}|d) \coloneqq pos_{\textsc{bbh}}(\vec{\theta}_\textsc{bbh})$ and $\pi(x') = p(k=\textsc{nsbh}, \vec{\theta}_\textsc{nsbh}|d) \coloneqq pos_{\textsc{nsbh}}(\vec{\theta}_{\textsc{nsbh}})$, and the deterministic mapping $(\vec{\theta}_{\textsc{bbh}, \vec{u}})$ = $h (\vec{\theta}_\textsc{nsbh}, \vec{v})$ and $(\vec{\theta}_\textsc{nsbh}, \vec{v}) = h^{-1} \vec{\theta}_{\textsc{bbh}, \vec{u}})$ are given by Eqs.~\eqref{eq:bbh_to_nsbh}-\eqref{eq:bbh_to_nsbh}, respectively. 
In the following, we explicitly write the model label as $pos_k$, instead of including it in the state as in $\pi(k,\vec{\theta}_k)$.
Therefore,
\begin{equation}
    pos_{\textsc{bbh}} \, g(\vec{u}) \, \alpha_{\textsc{bbh} \rightarrow \textsc{nsbh}} = pos_{\textsc{nsbh}} \, g'(\vec{v}) \, \alpha_{\textsc{nsbh} \rightarrow \textsc{bbh}} \, \mathbb{J}_{\textsc{bbh} \rightarrow \textsc{nsbh}},
\end{equation}
where $\mathbb{J}_{\textsc{bbh} \rightarrow \textsc{nsbh}}$ denotes the Jacobian relative to the mapping between the NSBH and BBH space, and for now we dropped the parameters' dependence in the posterior for simplicity.

Following the derivation of the expression for the acceptance ratio discussed in Sec.~\ref{ssec:mh}, we can define
\begin{equation}
    \alpha_{\textsc{bbh} \rightarrow \textsc{nsbh}} = \frac{s_{\textsc{bbh} \rightarrow \textsc{nsbh}}}{1 + \frac{pos_{\textsc{bbh}}g(u)}{pos_\textsc{nsbh}g'(v) \, \mathbb{J}_{\textsc{bbh} \rightarrow \textsc{nsbh}}}}
\end{equation}
and similarly for $\alpha_{\textsc{nsbh} \rightarrow \textsc{bbh}}$.
As before, the only constraint is that $s$ is symmetric, i.e., $s_{\textsc{bbh} \rightarrow \textsc{nsbh}} = s_{\textsc{nsbh} \rightarrow \textsc{bbh}}$. Hence, we can define
\begin{equation}
    s_{\textsc{bbh} \rightarrow \textsc{nsbh}} = \begin{cases}
     1 + \frac{pos_\textsc{bbh}g(u)}{pos_\textsc{nsbh}g'(v) \mathbb{J}}_{\textsc{bbh} \rightarrow \textsc{nsbh}} &\mbox{if} \; \; \frac{pos_\textsc{nsbh}g'(v)\mathbb{J}_{\textsc{bbh} \rightarrow \textsc{nsbh}}}{pos_\textsc{bbh}g(u)} \geq 1 \\
     1 + \frac{pos_\textsc{nsbh}g'(v) \mathbb{J}_{\textsc{bbh} \rightarrow \textsc{nsbh}}}{pos_\textsc{bbh}g(u)} &\mbox{if} \; \; \frac{pos_\textsc{nsbh}g'(v)\mathbb{J}_{\textsc{bbh} \rightarrow \textsc{nsbh}}}{pos_\textsc{bbh}g(u)} < 1, 
    \end{cases}
\end{equation}
which results in
\begin{equation}
    \alpha_{\textsc{bbh} \rightarrow \textsc{nsbh}} = \begin{cases}
        1 &\mbox{if} \; \; \frac{pos_\textsc{nsbh}g'(v)\mathbb{J}_{\textsc{bbh} \rightarrow \textsc{nsbh}}}{pos_\textsc{bbh}g(u)} \geq 1 \\
        \frac{pos_\textsc{nsbh} g'(v) \mathbb{J}_{\textsc{bbh} \rightarrow \textsc{nsbh}}}{pos_\textsc{bbh} g(u)} &\mbox{if} \; \; \frac{pos_\textsc{nsbh}g'(v)\mathbb{J}_{\textsc{bbh} \rightarrow \textsc{nsbh}}}{pos_\textsc{bbh}g(u)} < 1.
    \end{cases}
\end{equation}

Focusing on the case $\frac{pos_\textsc{nsbh}g'(v)\mathbb{J}_{\textsc{bbh} \rightarrow \textsc{nsbh}}}{pos_\textsc{bbh}g(u)} < 1$, since in the other one simply $\alpha_{\textsc{bbh} \rightarrow \textsc{nsbh}}=1$, we can use Bayes' theorem to write the acceptance probability as
\begin{equation}
    \alpha_{\textsc{bbh} \rightarrow \textsc{nsbh}} = \frac{pos_\textsc{nsbh} g'(v) \mathbb{J}_{\textsc{bbh} \rightarrow \textsc{nsbh}}}{pos_\textsc{bbh} g(u) } 
    = \frac{\mathcal{L}_\textsc{nsbh}(\vec{\theta}_\textsc{nsbh}, \Lambda_{1, \textsc{nsbh}}) p_\textsc{nsbh}(\vec{\theta}_\textsc{nsbh}, \Lambda_{1, \textsc{nsbh}}) g'(v) \mathbb{J}_{\textsc{bbh} \rightarrow \textsc{nsbh}}}{\mathcal{L}_\textsc{bbh}(\vec{\theta}_\textsc{bbh}, \Lambda_{1, \textsc{bbh}}, \Lambda_{2, \textsc{bbh}}) p_\textsc{bbh}(\vec{\theta}_\textsc{bbh}, \Lambda_{1, \textsc{bbh}}, \Lambda_{2, \textsc{bbh}})g(u)},
\end{equation}
where $\mathcal{L}$ denotes the likelihood and $p$ the prior.
Factorizing the prior we obtain
\begin{align}
    \alpha_{\textsc{bbh} \rightarrow \textsc{nsbh}} =
    &= \frac{\mathcal{L}_\textsc{nsbh}(\vec{\theta}_\textsc{nsbh}, \Lambda_{1, \textsc{nsbh}}) p_\textsc{nsbh}(\vec{\theta}_\textsc{nsbh}) p_\textsc{nsbh}(\Lambda_{1, \textsc{nsbh}}) g'(v) \mathbb{J}_{\textsc{bbh} \rightarrow \textsc{nsbh}}}{\mathcal{L}_\textsc{bbh}(\vec{\theta}_\textsc{bbh}, \Lambda_{1, \textsc{bbh}}, \Lambda_{2, \textsc{bbh}}) p_\textsc{bbh}(\vec{\theta}_\textsc{bbh}) p_\textsc{bbh}(\Lambda_{1, \textsc{bbh}}) p_\textsc{bbh}(\Lambda_{2, \textsc{bbh}})g(u)} \\
    &= \frac{\mathcal{L}_\textsc{nsbh}(\vec{\theta}_\textsc{nsbh}) p_\textsc{nsbh}(\vec{\theta}_\textsc{nsbh}) p_\textsc{nsbh}(\Lambda_{1, \textsc{nsbh}}) g'(v) \mathbb{J}_{\textsc{bbh} \rightarrow \textsc{nsbh}}}{\mathcal{L}_\textsc{bbh}(\vec{\theta}_\textsc{bbh}) p_\textsc{bbh}(\vec{\theta}_\textsc{bbh}) p_\textsc{bbh}(\Lambda_{1, \textsc{bbh}}) p_\textsc{bbh}(\Lambda_{2, \textsc{bbh}})g(u)},
    \label{eq:acc_prior}
\end{align}
where in the last step we used the fact that the likelihood does \emph{not} depend on the \textit{pseudo-}parameters.

Equation~\eqref{eq:acc_prior} shows how, even if the \textit{pseudo-} parameters do not enter the likelihood calculation, we have to include their prior in the acceptance ratio calculation in order to respect the detailed balance.

\section{Convergence criteria}
\label{sec:convergence_crit}

As discussed in Sec.~\ref{sec:convergence}, it is not trivial to establish a criterion to automatically check the convergence of RJMCMC analyses (see, e.g., Refs.~\cite{2010arXiv1001.2055F, 6fdd3944-b18c-3ac9-ab38-8f3f253b2638} for some examples and discussion). 
By default, \texttt{t-roo} runs for a set number of steps and, if needed, the run can be resumed and run for longer. 
However, we also implemented some ``empirical" convergence criteria that can be used to decide when to stop the sampler, both in the preliminary and main RJMCMC sampling phases. 
All the different stopping criteria available in \texttt{t-roo} are described below.

For the preliminary sampling stage, we implemented two stopping criteria:
\begin{itemize}
    \item \textit{Number of steps} (default): the preliminary sampling is stopped after a given number of steps, defined by the user. Although different analyses might require a different number of steps, we find that in general $\mathcal{O}(10^3)$ is sufficient and not too computationally expensive. 
    \item \textit{Likelihood convergence}: the preliminary sampling is stopped when $80 \%$ of the walkers have converged to the maximum likelihood among them within a given tolerance, where the default tolerance is set to $0.1$. The idea behind this criterion is that when the sampler finishes exploring the parameter space and finds the region(s) where the support for each model lives, the likelihood of the samples around that region will stabilize around the maximum value and yield similar values among the various walkers. 
\end{itemize}

For the main RJMCMC sampling, we explored three different stopping criteria. In this case, we do not look at the likelihood but at the model probabilities, because in RJMCMC the number of samples in each model should converge to the respective probabilities of that model. The underlying idea is that, once the sampler is converged, each walker will ideally spend an amount of time (i.e., samples) in each model corresponding to that model probability:
\begin{itemize}
    \item \textit{Number of steps} (default): the run is stopped after a specified number of iterations.
    \item \textit{Walkers mean}: we compute the average probability among walkers of each model in the last $n$ steps, and we stop the run when this average does not change for a given number of steps.
    \item \textit{Multinomial}: the run is stopped when the standard deviation between each model probability as computed from different walkers is less than 1.3 times the standard deviation expected from a multinomial distribution for $N_w$ walkers.
\end{itemize}
In order to save computational time, the users can decide after how many iterations to start checking whether stopping criteria are satisfied, and how often to check after. 
In any case, the run can always be resumed and run for more steps.

\section{Robustness of evidence estimation}
\label{sec:evidences}

\begin{figure}[htb]
        \centering
        \includegraphics[width=0.6\linewidth]{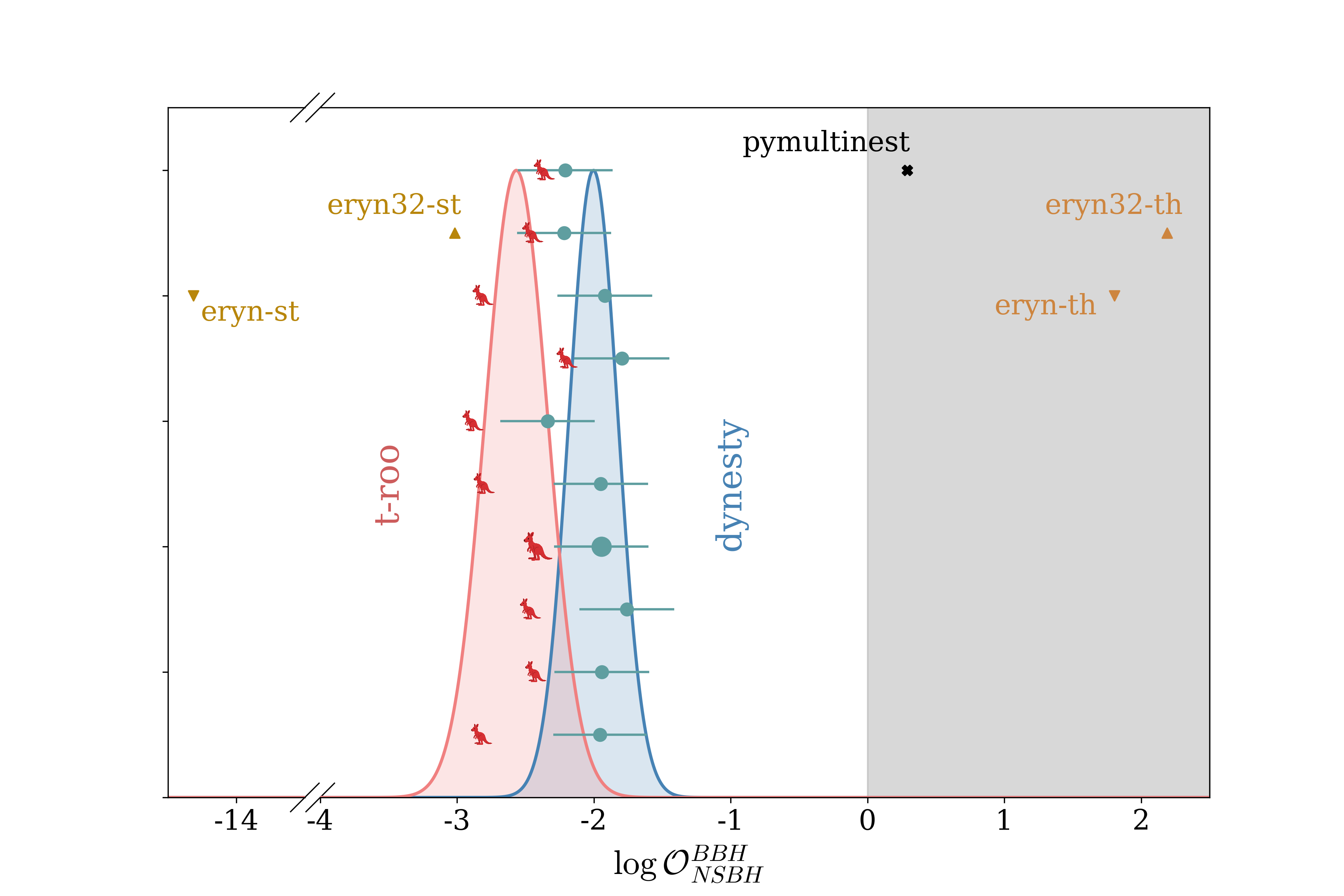}
    \caption{Comparison between odds ratios for the \textsc{nsbh\_lam600} run with different samplers. 
    The gray shaded area marks the region in which the BBH model is incorrectly favored. The blue and red Gaussians show the expected distributions of odds ratio from \texttt{dynesty} and \texttt{t-roo}, respectively, as inferred from the results of 10 analyses with different sampling seeds. The markers show the $\ln \mathcal{O}_{\textsc{nsbh}}^{\textsc{bbh}}$ recovered by each run, with the larger marker highlighting the run whose results are reported in Table~\ref{tab:results_comp}; in the case of \texttt{dynesty}, the error bar corresponds to the conservative estimate of the uncertainty as described in Sec.~\ref{sec:results}. For \texttt{eryn} runs, ``th" and ``st", refer to evidences computed via the thermodynamic integration and stepping stone methods, respectively, while ``eryn-32" refers to the \texttt{eryn} runs with 32 temperatures. }
    \label{fig:evidence_comp}
\end{figure}

Throughout this work, we have compared the \texttt{t-roo} model probabilities to the ones obtained by the default nested sampler in \textsc{bilby}, i.e., \texttt{dynesty}.
In order to check the robustness of the evidence calculation, we here compare its results to evidences obtained by different methods, focusing on the injection \textsc{nsbh\_lam600}; results are shown in Fig.~\ref{fig:evidence_comp}.

We repeat the \texttt{dynesty} analysis for the \textsc{nsbh\_lam600} injection with different sampling seeds, using 2000 live points in all cases\footnote{Ref.~\cite{Romero-Shaw:2020owr} showed in Sec.~A2 that the true evidences consistently lie within the errors provided by \texttt{dynesty} 68\% of the times when more than 1000 live points are used, even though the error on \texttt{dynesty} evidences is not technically Gaussian.}. 
The odds ratio estimate within \texttt{dynesty} is robust, with the fluctuations due to different sampling seeds lying within the uncertainty band, therefore it constitutes an appropriate reference to assess the reliability of the sampler we developed. 

Similarly, we repeat the \texttt{t-roo} analysis with different sampling seeds: also in this case we see fluctuations in the recovered odds ratio value, but the estimate is overall consistent across the different sampling seeds.
The blue and red Gaussians in Fig.~\ref{fig:evidence_comp} represent the distribution of odds ratio values from the \texttt{dynesty} and \texttt{t-roo} analyses of this injection, respectively, with mean and standard deviation derived from the distribution of 10 runs with different sampling seeds. 

Furthermore, we also analyze this injection with a different nested sampler, i.e., \texttt{pymultinest}~\cite{Feroz:2008xx, Buchner:2014nha}. In this case, the recovered $\ln \mathcal{O}_{\textsc{nsbh}}^{\textsc{bbh}}$ deviates substantially from the ones recovered by \texttt{dynesty}, even favoring the BBH model instead of the NSBH one.

Since \texttt{t-roo} is based on \texttt{eryn}, we finally check the evidences obtained by single-model \texttt{eryn} runs, without the specific within- and between- model moves described above. 
MCMC algorithms sample directly the target posterior distribution, and consequently do not automatically provide Bayesian evidences. 
However, one can exploit the parallel tempering chains to compute them. 
In particular, two different methods are implemented in \texttt{eryn}: \textit{thermodynamic integration}~\cite{thermo_int} and \textit{stepping stone}~\cite{st_st}. 
Given that the accuracy in evidence estimates increases with the number of parallel temperature chains employed, we additionally repeated these runs with 32 temperatures instead of 16. 
The odds ratios obtained with \texttt{eryn} show large differences not only with respect to the \texttt{dynesty} runs, lying well outside the uncertainty region and, in the case of thermodynamic integration, even favoring the wrong model. 
However, it is known that accurate thermodynamic evidences require a sufficiently large number of temperatures, and that evidences computed with the stepping stone algorithm require fewer temperatures than standard thermodynamic integration. Therefore, we expect the \texttt{eryn32-st} evidences to be most accurate, and this is indeed the run that agrees with the \texttt{t-roo} evidence. Most likely the inaccuracies in the other odds ratio estimations are due to insufficient temperatures. This shows another advantage of \texttt{t-roo}, meaning that it can obtain good estimates of odds ratios with less temperatures (i.e., further reducing the computational cost of the analysis).

Overall, Fig.~\ref{fig:evidence_comp} shows how different methods can provide quite different estimates of evidence values, and consequently odds ratios, calling for general caution when basing strong claims on them. 
Nevertheless, the odds ratio estimation in \texttt{t-roo} appears to be quite robust, with minimum fluctuations across the results for different sampling seeds.

\twocolumngrid 

\bibliography{bibliography}{}
\bibliographystyle{apsrev4-1}

\end{document}